\pdfoutput=1
\documentclass[traditabstract]{aa} 
\usepackage{graphicx}
\usepackage{txfonts}
\usepackage{color}

\begin{document}

   \title{The dense molecular gas in the $\rm z\sim6$ QSO \\ 
   SDSS J231038.88+185519.7 resolved by ALMA}

   \author{C. Feruglio
          \inst{1}
          \and
          F. Fiore \inst{1}
        \and S. Carniani \inst{2,3} 
        \and R. Maiolino \inst{2,3}     
        \and V. D'Odorico \inst{1}
        \and  A. Luminari \inst{4,5}
        \and P. Barai \inst{6}
        \and M. Bischetti \inst{4,5}
        \and A. Bongiorno \inst{5}
        \and  S. Cristiani \inst{1}
        \and A. Ferrara \inst{7}
        \and S. Gallerani \inst{7}
        \and A. Marconi \inst{8,9}
        \and A. Pallottini \inst{7,10}
        \and E. Piconcelli \inst{5}
        \and L. Zappacosta \inst{5}        
}

   \institute{INAF Osservatorio Astronomico di Trieste, Via G. Tiepolo 11, Trieste, Italy \\
              \email{chiara.feruglio@inaf.it}
         \and Cavendish Laboratory, University of Cambridge, 19 J. J. Thomson Ave., Cambridge CB3 0HE, UK
         \and Kavli Institute for Cosmology, University of Cambridge, Madingley Road, Cambridge CB3 0HA, UK
         \and Dipartimento di Fisica, Universit\'a di Roma "Tor Vergata", Via della Ricerca Scientifica 1, 00133 Roma, Italy 
         \and INAF Osservatorio Astronomico di Roma, via Frascati 33, 00078 Monte Porzio Catone, Italy
        \and Instituto de Astronomia, Geof\'isica e Ci\^encias Atmosf\'ericas - Universidade de S\~ao Paulo (IAG-USP), 
                Rua do Mat\~ao 1226, S\~ao Paulo, 05508-090, Brasil 
        \and Scuola Normale Superiore, Piazza dei Cavalieri 7, 56126 Pisa, Italy 
         \and Dipartimento di Fisica e Astronomia, Universit\'a di Firenze, via G. Sansone 1, I-50019, Sesto Fiorentino (Firenze), Italy
        \and INAF-Osservatorio Astrosico di Arcetri, Largo E. Fermi 2, I-50125, Firenze, Italy   
         \and Centro Fermi, Museo Storico della Fisica e Centro Studi e Ricerche "Enrico Fermi", Piazza del Viminale 1, I-00184 Roma, Italy
         }
             
   \date{}
 
  \abstract
{We present ALMA observations of the CO(6-5) and [CII] emission lines and the sub-millimeter continuum of the $z\sim6$ quasi-stellar object (QSO) SDSS J231038.88+185519.7. Compared to previous studies, we have analyzed a synthetic beam that is ten times smaller in angular size, we have achieved ten times better sensitivity in the CO(6-5) line, and two and half times better sensitivity in the [CII] line, enabling us to resolve the molecular gas emission.
We obtain a size of the dense molecular gas of $2.9\pm0.5$ kpc, and of $1.4\pm0.2$ kpc for the 91.5 GHz dust  continuum. By assuming that CO(6-5) is thermalized, and by adopting a CO--to--$H_2$ conversion factor $\rm \alpha_{CO} = 0.8~ M_{\odot}~K^{-1}~ (km/s)^{-1} ~pc^{2}$, we infer a molecular gas mass of $\rm M(H_2)=(3.2 \pm0.2) \times 10^{10}\rm M_{\odot}$.
Assuming that the observed CO velocity gradient is due to an inclined rotating disk, we derive a dynamical mass of $\rm M_{dyn}~sin^2(i) = (2.4\pm0.5) \times 10^{10}~ M_{\odot}$, which is a factor of approximately two smaller than the previously reported estimate based on [CII]. 
Regarding the central black hole, we provide a new estimate of the black hole mass based on the C~IV emission line detected in the X-SHOOTER/VLT spectrum: $\rm M_{BH}=(1.8\pm 0.5)  \times 10^{9}~ M_{\odot}$. 
We find a molecular gas fraction of $\rm \mu=M(H_2)/M^*\sim4.4$, where $\rm M^*\approx M_{dyn}  - M(H_2)-M(BH)$.
We derive a ratio $v_{rot}/\sigma \approx 1-2$ suggesting high gas turbulence, outflows/inflows and/or complex kinematics due to a merger event. 
We estimate a global Toomre parameter $Q\sim 0.2-0.5$, indicating likely cloud fragmentation.
We compare, at the same angular resolution, the CO(6-5) and [CII] distributions, finding that dense molecular gas is more centrally concentrated with respect to [CII].
We find that the current BH growth rate is similar to that of its host galaxy. 
  }

   \keywords{ (galaxies): quasars: individual : SDSS J231038.88+185519.7 -- galaxies: high-redshift -- galaxies: ISM -- (galaxies): quasars: general -- (galaxies): quasars: supermassive black holes
               }

   \maketitle
%

\section{Introduction}

To date, about 200 quasi-stellar objects (QSOs) at z$\gtrsim 6$ are known, offering the opportunity to study the early growth and co-evolution of super massive black holes (SMBH) and their host galaxies.  
These rare systems are the sites where large $10^{11.5} -10^{12}~M_{\odot}$ stellar masses are predicted to aggregate, and thus they are the likely cradles of local giant galaxies.
Optical and near-infrared (NIR) observations of these high-z QSOs lead to measured black hole masses often exceeding $10^9\rm~M_{\odot}$ (Jiang et al. 2007, De Rosa et al. 2014, Venemans et al. 2015,  Ba\~nados, 2016, 2018). 
Sub-millimeter observations revealed compact rotating disks, active star formation in their host galaxies, and large masses of highly excited molecular gas. Optical/NIR observations have led to estimations of SMBH-stellar mass ratios that are larger than the local universe
value by factors of between a few and ten (Walter et al. 2004, Maiolino et al. 2005, 2009, Carilli et al. 2007, Wang et al. 2010, 2013, 2016, Gallerani et al. 2014, Willott et al. 2015, Venemans et al. 2016, 2017ab, Decarli 2017, 2018).  
Both simulations and semi-analytical models of galaxy formation showed that SMBHs can grow either from pop III star seeds at super-Eddington rates, or from massive seeds, and assemble BH masses of several times $10^9\rm ~ M_{\odot}$ observed at $z\gtrsim 6$ (Volonteri et al. 2016, Valiante et al. 2016, Pezzulli et al. 2016, 2017).  

The [CII]~158 $\mu$m emission line is the preferred tracer used to study the host galaxies of high-z QSOs because its brightness makes it a powerful probe to survey large samples by snapshot observations (Decarli et al. 2018).  
Conversely, the dense molecular ISM, where we expect most star formation to take place, remains relatively poorly explored at these high redshifts (Wang et al. 2010, 2013, Gallerani et al. 2014, Venemans et al. 2017a).
Finally, the host galaxies of high-z QSOs are mostly spatially unresolved in current [CII] and CO observations, making the estimate of dynamical masses very uncertain.   
Today it is possible to resolve the molecular gas emission in the highest-redshift QSO host galaxies with the Atacama Large Millimeter/submillimeter Array (ALMA). 

This work focuses on the QSO J231038.88+185519.7 (hereafter J2310+1855), the QSO with the brightest 250 GHz continuum at z$\sim$6 (8.29 mJy, Wang et al. 2013, hereafter W13).  
Using the Northern Extended Millimeter Array (NOEMA) W13 detected the CO(6-5) emission line from the host galaxy of the QSO and measured an emission redshift of $z=6.0025$. 
W13 also detected the [CII] emission line, and marginally resolved the rotation of a disk, from which they derived a dynamical mass of $M_{dyn}= (9.6\pm0.6) \times 10^{10}~\rm M_{\odot}$, and an inclination of 46 deg, estimated from major/minor axis ratio. 

Jiang et al. (2016) detected weak C~IV and Mg~II emission lines in the GEMINI/GNIRS spectrum, from which they derived a redshift of $5.962\pm0.007$,  which implies that the UV lines are blueshifted by $\sim 1200$ km/s with respect to CO and [CII] lines, similarly to what was found for other $z>6$ QSOs (Willott et al. 2015, Venemans et al. 2016, Wang et al. 2017), and common also at lower redshift
(Tytler  \& Fan 1992). 
They derived a BH mass of $(3.9\pm0.5)-(4.2\pm1.0)~\times 10^{9}\rm M_{\odot}$, based on Mg~II and C~IV, respectively.

In this work, we report on our ALMA observations of CO(6-5), the sub-mm continuum (Project 2015.1.00584.S), the [CII] emission line (Project 2015.1.00997.S) of the J2310+1855 host galaxy, and on archival X-SHOOTER/VLT (Programme 098.B-0537(A), P.I. Farina) observations of the rest-frame ultraviolet (UV)-optical spectrum of the QSO.  
Compared to W13, we have used a synthetic
beam that is ten times smaller in angular size, and we have achieved ten times
better sensitivity in the CO(6-5) line, and two and half times better sensitivity
in the [CII] line, enabling us to resolve the molecular gas emission of the QSO host galaxy.
The observations are described in Section 2. 
In section 3 we present results on CO and [CII] kinematics and their ratio, gas mass, dynamical mass, black hole mass, and detection of other line-emitting sources. 
In Section 4 we discuss the results, and a summary is presented in Section 5. 
A $\Lambda$-CDM cosmology with $H_0=70$ km s$^{-1}$ Mpc$^{-1}$, $\Omega_M=0.3$, and $\Omega_\Lambda=0.7$ is adopted throughout the paper. The angular scale is 5.835 kpc/\arcsec~ for the adopted cosmology.



\section{Observations}

\subsection{ALMA observations}

We observed J2310+1855 with ALMA band-3 receivers tuned to cover the frequency ranges 84.56-87.94, and 96.56-99.69 GHz.
Spectral window 1 was tuned at the expected redshifted frequency of CO(6-5), 98.75 GHz. 
We performed calibration in the CASA environment (McMullin et al. 2007).
Mapping and data analysis were performed both in the CASA and in the GILDAS (Guilloteau et al. 2000) environments (the latter after converting CASA into GILDAS visibility tables).  
Within GILDAS, we created two data cubes, one per baseband. We also created a continuum map using the four spectral windows and excluding channels within $\pm 1000$ km/s from the redshifted CO(6-5) frequency.
The continuum map and the data cubes were cleaned by adopting a natural weight scheme, with detection threshold equal to 0.5 times the noise (per channel), and without applying any mask for the detection. We  used both the {\it hogbom} and the {\it mx} cleaning algorithms within GILDAS, and found consistent results. 
 In the following, we use the results from the {\it hogbom} cleaning, because it minimizes side-lobe residuals. Based on this setup we obtain a synthesized beam of $0.6 \times 0.4$ arcsec$^2$ at a PA$=-6$ deg in the cubes, which contains the CO(6-5) emission line.
The noise levels are 5.4 $\mu$Jy/beam in the continuum in the aggregated bandwidth, and 0.13 mJy/beam in the 23.7 km/s-wide channels (i.e., the maximum spectral resolution of the data).
By adopting a Briggs algorithm, the synthetic beam is $0.51 \times 0.28$ arcsec, at a PA$=-11$ deg, 
and the noise level is 0.15 mJy/beam in the 23.7 km/s-wide channel. 

[CII] observations were obtained with the ALMA 12-m array in  project 2015.1.00997.S.  
The data were calibrated and imaged in CASA v4.7 by applying a natural weighting with a detection threshold equal to 0.5 times the noise (per channel). The continuum-subtracted cube was obtained by combining two adjacent spectral windows after subtracting the continuum emission by fitting a UV-plane model with the {\it uvcontsub} task. The $1\sigma$ root mean square (r.m.s) sensitivity is 0.20 mJy/beam per 100 km/s channel. The synthesized beam is $0.9\times0.6$ arcsec$^2$ at a PA=49 deg.

In this paper we aim at comparing the kinematics and spatial distribution of CO- and [CII]-emitting gas.  
A thorough description of [CII] luminosity, mass, and other host galaxy properties inferred from [CII] are the subject of a separate paper (Carniani et al. 2018 in prep.). 
In order to compare CO and [CII] emissions, we imaged the CO(6-5) data with the same restoring beam as the [CII] data, and we registered the two cubes at the same reference system, that is, z$=6.0025$ in the LSRK reference frame, with the same spectral binning (23.7 km/s spectral resolution).


\subsection{X-SHOOTER spectrum}

We used archival data from X-SHOOTER at the VLT (Vernet et al. 2011) to investigate the optical/NIR spectrum of the QSO.
There are two frames available in the archive with an exposure time of 1200 s each, observed with a slit
of $0.9$ arcsec and a binning $2\times2$ in the VIS arm (550-1020 nm) and a slit of
$0.6$ arcsec in the NIR arm (1020-2480 nm). 
The chosen slits correspond to nominal resolving powers of $R\simeq 8800$ and $8100$, respectively.

The spectra were reduced with the ESO pipeline (Modigliani et
al. 2010) with a manual localisation of the object, and adopting the sky subtraction
method BSPLINE1 in the VIS, and MEDIAN in the NIR. The one-dimensional (1D) flux
calibrated spectra produced by the pipeline were then corrected for
telluric absorption using the ESO tool {\tt Molecfit} (Smette et al. 2015, Kausch et al. 2015). 
The final spectra, obtained combining the two frames, 
were rebinned to a step of 0.4 and 0.6 \AA, respectively.
In order to fit the emission lines (see section 3.4), the NIR spectrum was then binned to a step of 7 \AA.

\section{Results}

\subsection{The 3.3 mm continuum}

Regarding the 91.5 GHz ($\sim 3.3$ mm) continuum, 
we performed a fit of the continuum visibilities in both CASA and GILDAS. Both methods give consistent results and show that the continuum source is best fitted by a 2D Gaussian function with parameters reported in Table 1 (errors do not include the systematic error on the flux scale, which is of the order $\approx 5\%$ at this frequency).
The 91.5 GHz continuum flux density is $416\pm33~ \rm \mu Jy$, consistent with the measurement by W13 at a similar frequency. 
The continuum has a beam--deconvolved FWHM size of $(0.24\pm0.04)\times(0.12\pm0.07)$ arcsec$^2$, at a PA=$138\pm24$ deg, which corresponds to a physical size of $1.4\pm0.2$ kpc (Table 1).  
This measurement is in agreement with the continuum size measured by W13 at 263 GHz frequency (the 3 mm continuum is unresolved in the PdBI observations presented in W13).


\begin{table}
\caption{91.5 GHz continuum, CO(6-5) and [CII] best fit parameters.} 
\label{table:1}      
\centering                          
\begin{tabular}{l l r}        
\hline
\hline
\smallskip
RA$_{\rm cont,91.5GHz}$  &    23:10:38.90 $\pm$ 0.02 & [J2000] \\
\smallskip
DEC$_{\rm cont,91.5GHz}$  & 18:55:19.82 $\pm$ 0.02 & [J2000] \\
\smallskip
Size$_{\rm cont,91.5GHz}$ &  $(0.24\pm0.04)\times(0.12\pm0.07)$ & [arcsec$^2$] \\
\smallskip
PA$_{\rm cont,91.5GHz}$ & $138\pm24$ & [deg] \\ 
\smallskip
$S_{91.5GHz}$  &  $416\pm33$ & $ \rm  [\mu Jy]$  \\ 
\hline
\smallskip
RA$_{\rm CO(6-5)}$  & 23:10:38.900 $\pm$ 0.008 & [J2000] \\
\smallskip
DEC$_{\rm CO(6-5)}$  & 18:55:19.83 $\pm$ 0.01  &[J2000]\\
\smallskip
Size$_{\rm CO(6-5)}$ & $(0.33\pm0.06) \times (0.20\pm0.04)$ & [arcsec$^2$]\\ 
\smallskip
PA$_{\rm CO(6-5)}$ &  $140\pm17$ & [deg]\\
\smallskip
$Sdv_{\rm CO(6-5)}$  & $1.26\pm0.06$ & [Jy km/s] \\ 
\hline
\smallskip
RA$_{\rm [CII]}$  & 23:10:38.929 $\pm0.09$  & [J2000] \\
\smallskip
DEC$_{\rm [CII]}$  & 18:55:18.1 $\pm0.1$   &[J2000]\\
\smallskip
Size$_{\rm [CII]}$ & $(0.86\pm0.10)\times(0.78\pm0.15)$  & [arcsec$^2$]\\ 
\smallskip
PA$_{\rm [CII]}$ &  $147\pm68$ & [deg]\\
\smallskip
$Sdv_{\rm [CII]}$  & $8.75\pm0.51$ & [Jy km/s] \\ 
\hline
\hline
\end{tabular}
\end{table}

\subsection{CO(6-5) emission line}

We created a continuum-subtracted data cube by using the task {\it uvcontsub} in CASA. 
To this aim, we combined spectral windows 1 and 2 (so as to have a broader bandwidth around the emission line for continuum estimate). The continuum was then estimated by fitting the line-free channels (at velocity $>|1000| ~\rm km/s$ from the line peak) with both a constant and a first-order polynomial. By eye, no slope is seen in the continuum, and we verified that a zero-order fit gives the best result. 
The continuum-subtracted visibility set was then used to map the CO(6-5) emission and to create a visibility set averaged over the line channels, in the velocity range from $-500$ to 500 km/s from the systemic velocity. 

We used the velocity-integrated visibility set to measure the line parameters in the {\it uv} plane. We fitted the visibilities with a point source and a circular and an elliptical Gaussian function. The best fit is given by the elliptical Gaussian function which delivers a source size (FWHM) of $(0.33\pm0.06) \times (0.20\pm 0.04)$ arcsec, at a $\rm PA=140\pm 17$ deg (Table \ref{table:1}). 
This size is smaller than that derived for [CII] by W13, suggesting that the CO emitting gas is more centrally concentrated with respect to [CII].
The integrated flux density of the line is $1.26\pm0.06$ Jy km/s (consistent within 2$\sigma$ with W13).
The [RA, DEC] of the CO peak emission are consistent with the continuum position (Table \ref{table:1}).

\begin{figure*}
   \centering
  \includegraphics[width=\textwidth]{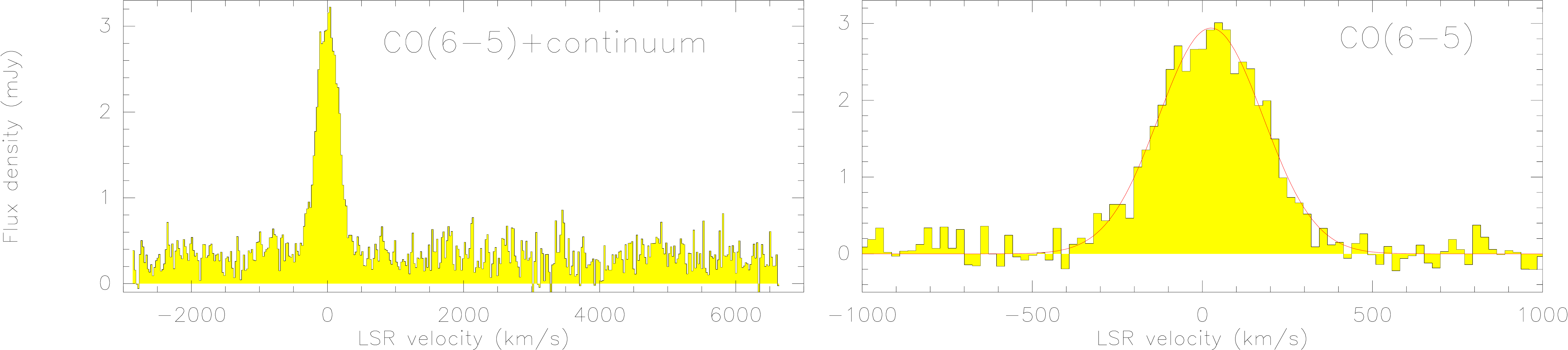}
  \caption{[Left panel]: Spectrum of spectral windows 1+2, including the CO(6-5) emission line and the continuum emission, extracted from the region included within the $\geq 2\sigma $ in the velocity-integrated map. [Right panel]: Zoom onto the continuum-subtracted CO(6-5) spectrum. The red line shows a fit with a Gaussian function. Spectra are plotted at the maximum spectral resolution of the data (23.7 km/s).}
              \label{Fig-sp}
    \end{figure*}

We then extracted the spectrum in Fig. \ref{Fig-sp} using a mask that encompasses the $\geq2\sigma$ region in the velocity-integrated CO(6-5) map.  
The CO(6-5) line peaks at a frequency corresponding to $z_{CO}=6.0028\pm0.0003$, consistent with previous CO and [CII]-based redshifts (W13). 
A Gaussian fit with a single component gives a peak intensity of the line of $I=2.94\pm0.06$ mJy, 
and FWHM$=361\pm9$ km/s. This gives an integrated flux density of $S_{CO}dv = 1.13\pm 0.06$ Jy km/s -- consistent with the estimate from the {\it uv fit} within $2\sigma$.

\begin{figure*}
   \centering
  \includegraphics[width=\textwidth]{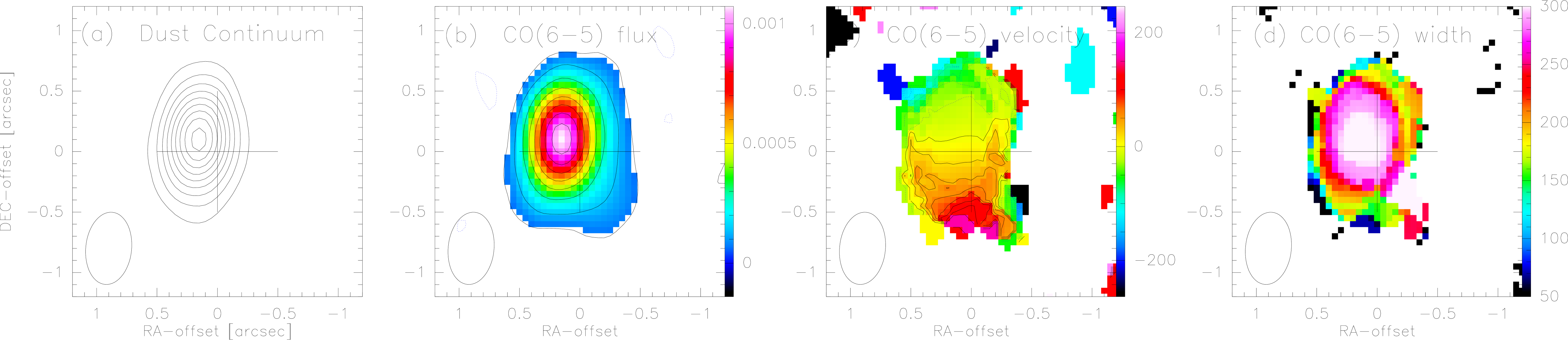}
  \caption{From left to right: The dust continuum map of J2310+1855 (levels from 2 to 45$\sigma$ by 5$\sigma$, $\sigma=5.36 ~\mu $Jy/beam), the moment 0, 1, 2 maps of the CO(6-5) emission line. Color-scale units are mJy and km/s, respectively. The cross indicates the phase center. The synthesized beam is indicated in the lower-left part of the diagrams.}
              \label{moments-co}
    \end{figure*}

Figure \ref{moments-co} shows the continuum map, and the maps of the first three moments of CO(6-5).  
The velocity map indicates a velocity gradient  of $\approx 300$ km/s from north-east to south-west, at a $\rm PA\approx 60$ deg north of east, consistent with that estimated for [CII] (see Fig. 4, and W13). 
The velocity dispersion map shows $\sigma \sim 300$ km/s at the center, which is likely affected by disk beam smearing effects. The intrinsic velocity dispersion may be significantly lower. According to Tacconi et al. (2013) and Davies et al. (2011) a less biased estimate of the velocity dispersion is given by the average value measured in the outer parts of the galaxy, assuming a flat profile. We then estimate $\sigma \approx 150-200$ km/s, similar to the value estimated for the [CII] (Fig. 4).
Under the assumption that the velocity gradient is due to a rotating disk, and based on the axis ratio (Table \ref{table:1}), we derive an inclination of the disk of $i\rm= cos^{-1}~ (minor/major) = 53$ deg (the lower limit inclination is $i_{min}\approx 25$ deg, by accounting for the statistical errors in minor/major axis). 
The rotation velocity of the disk would then be $v_{rot} = 1.3 \times \Delta v / 2~sin(i) \sim 245$ km/s (e.g., Tacconi et al. 2013). 
This is of the same order of the velocity dispersion given above, and in fact the ratio of the rotational velocity to the velocity dispersion, $v_{rot}/\sigma$, is in the range 1-2.
The large uncertainty on $v_{rot}/\sigma$ is driven by the uncertainty on the inclination and on the intrinsic unbiased $\sigma$.

\begin{figure*}
   \centering
    \includegraphics[width=8cm]{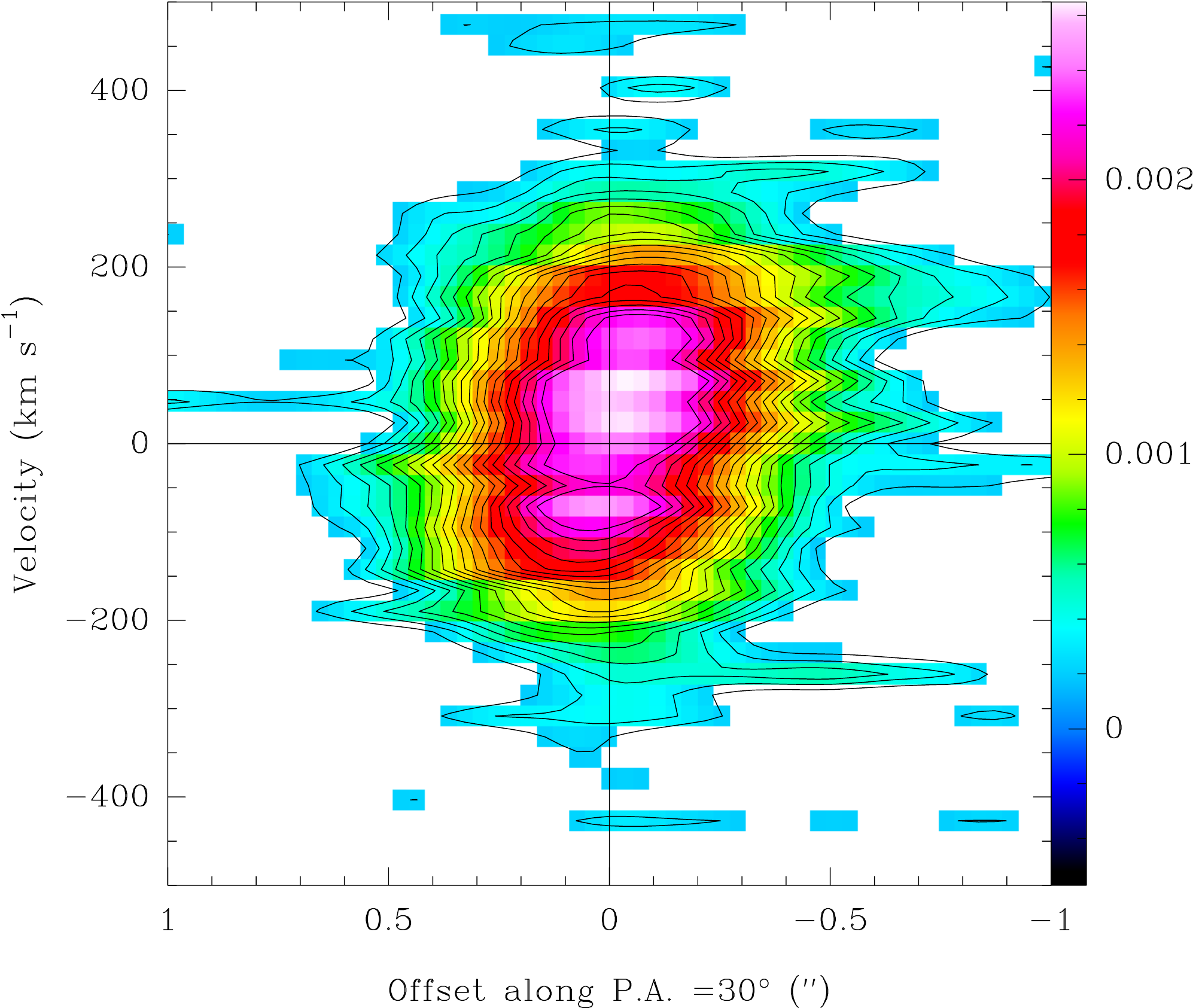}
    \includegraphics[width=8cm]{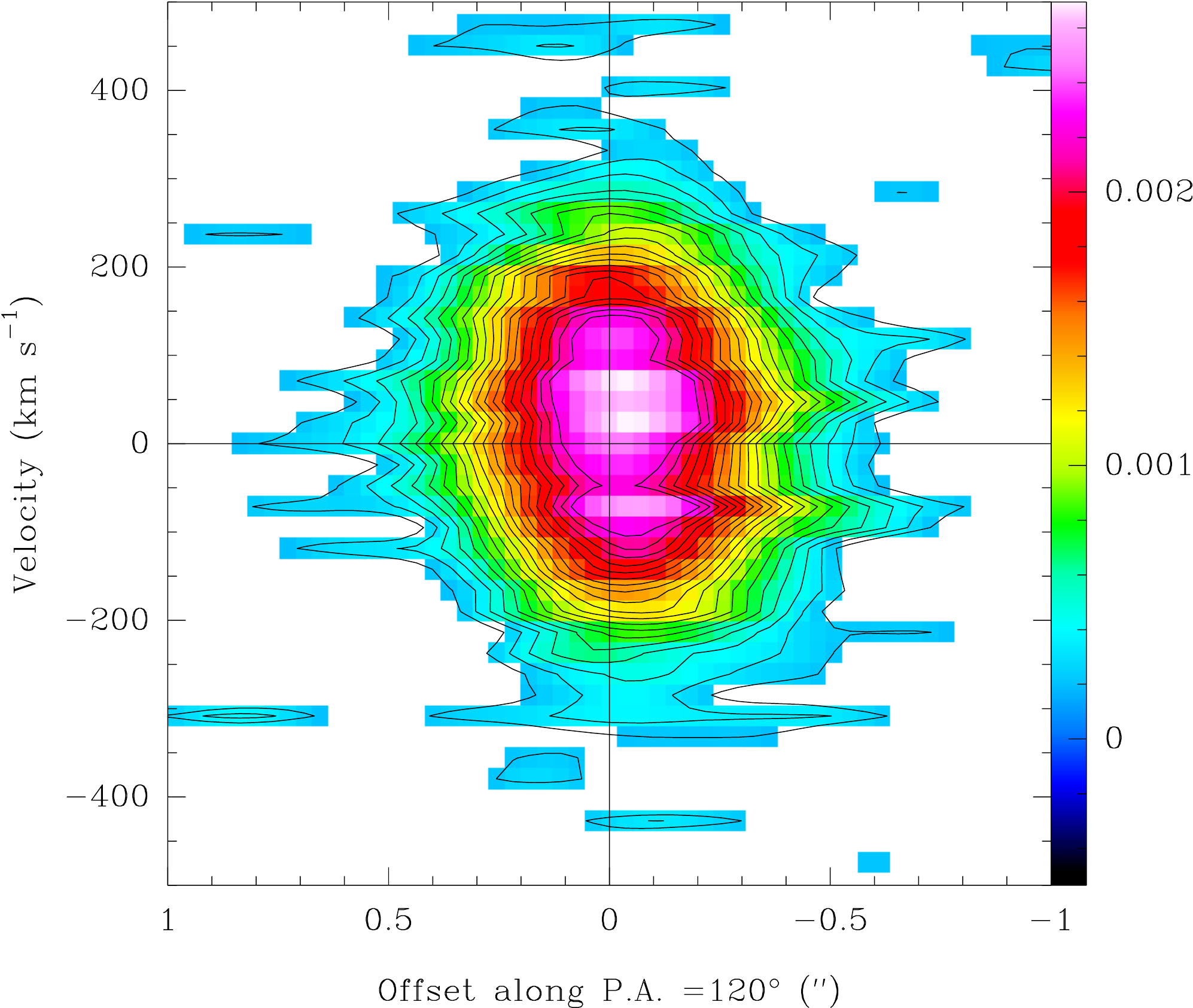}
  
  \caption{PV diagrams of the continuum-subtracted CO(6-5) emission line along the line of nodes (PA= 30 deg), and orthogonal to it (PA=120 deg). Contours are 2 to 20$\sigma$, by $\sigma$, $\sigma= 0.11$ mJy/beam. Slices are extracted from a slit of 0.5 arcsec in width. 
 Zero offset position is (RA, DEC)=(23:10:38.90 ,18:55:19.82) in these and the following PV diagrams.}
              \label{pv-co}
    \end{figure*}

Figure \ref{pv-co} shows position-velocity (PV) diagrams obtained along $\rm PA=30$ $\deg$ (i.e., along the velocity gradient) and orthogonal to it, through the peak CO position, with a slit of 0.5 arcsec in width. 
Non-rotational motions are visible in Fig. \ref{pv-co}, at velocities $+200$ km/s and -300 km/s at an offset of 0.7-0.5 arcsec on the south-west side of the QSO.
The emission from this relatively high-velocity gas is seen as an elongated structure located south-west of the nucleus in the CO(6-5) moment maps (Fig. 2).


\subsection{[CII] and the CO(6-5)/[CII] ratio}

\begin{figure*}
   \centering
  \includegraphics[width=\textwidth]{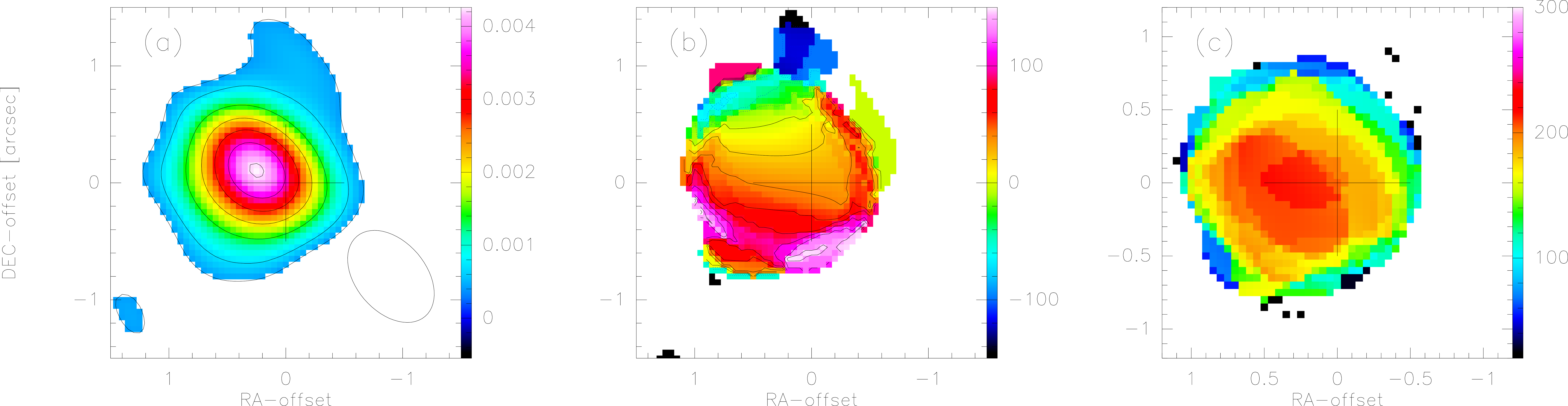}
  \caption{Moments of the [CII] emission line. (a) mean flux (levels are $-3,~ 3,~5,~ 10,~ 15,~ 20,~ 25,~ 30,~ 35\sigma,~\sigma=0.14 \rm ~ mJy/beam$); (b) velocity map; (c) velocity dispersion map. The synthesized beam is indicated in the left panel. The cross indicates the phase center.}
              \label{moments-cii}
    \end{figure*}

Figure \ref{moments-cii} shows the moment 0, 1, 2 maps of the [CII] emission line. 
The [CII] velocity gradient is found approximately along the same direction as the CO one. 
Figure 5 shows the PV diagram of the [CII] emission along the same directions as done for CO(6-5), with a larger slit width motivated by the larger beam of [CII] data. 
This shows that [CII] gas kinematics are similar to those of the molecular gas traced by CO(6-5).
We computed the ratio of CO(6-5)/[CII] by combining the respective velocity-integrated data cubes. Figure \ref{ratio} shows the velocity-integrated [CII] map (range $\pm500$ km/s), the map of CO(6-5) obtained by degrading the CO data to the same resolution as [CII] data, and their ratio. 
The CO(6-5)/[CII] ratio ranges from $\approx 0.1$ to 0.25, with a velocity-averaged mean ratio of 0.19.
We find that the CO(6-5)/[CII] ratio shows a local maximum close to the QSO position (indicated by the small black cross), and decreases at larger distances from the QSO (Fig. \ref{ratio}, right panel). 
This suggests that the CO(6-5)-emitting gas may be more concentrated than the [CII]-emitting gas. 

An excess in the CO(6-5)/[CII] ratio is seen  at a spatial offset of $(-0.3,-0.5)$ arcsec 
from the QSO (Fig. 6), at the position where a marginal disturbance is seen in both the velocity and velocity dispersion maps (Fig. 2). 

We extracted  the CO(6-5) and [CII] spectra from the respective data cubes, over a region centered at the QSO position (RA, DEC)=(23:10:38.90, 18:55:19.82), and with a radius of 0.5 arcsec for both lines, and compute their ratio (Fig. \ref{ratio-sp}).
We find that the line profiles are similar, although our data hint
at a broader velocity distribution of the [CII]-emitting gas compared to the CO(6-5)-emitting gas, within $\pm250$ km/s from the CO line peak.

Figure 8 shows PV diagrams of the CO(6-5)/[CII] ratio, obtained after combining the data cubes registered as detailed in Section 2.1. For consistency with the previous sections, we plot PV diagrams along the same PA used in the individual CO and [CII] diagrams. 
The PV diagrams are consistent with the global trend of the CO(6-5)/[CII] ratio found in the spectra, and show an average value of $\approx 0.2$. 
The PV diagrams suggest that the largest ratios are measured at a spatial offset of $(-0.3,-0.5)$ arcsec 
from the QSO, that is, at the position where we also find a local maximum of the $\sigma$ (Fig. 2). 

Confirmation of whether the small discrepancies in the line profiles are genuine would require additional observations.

  \begin{figure*}
   \centering
    \includegraphics[width=8cm]{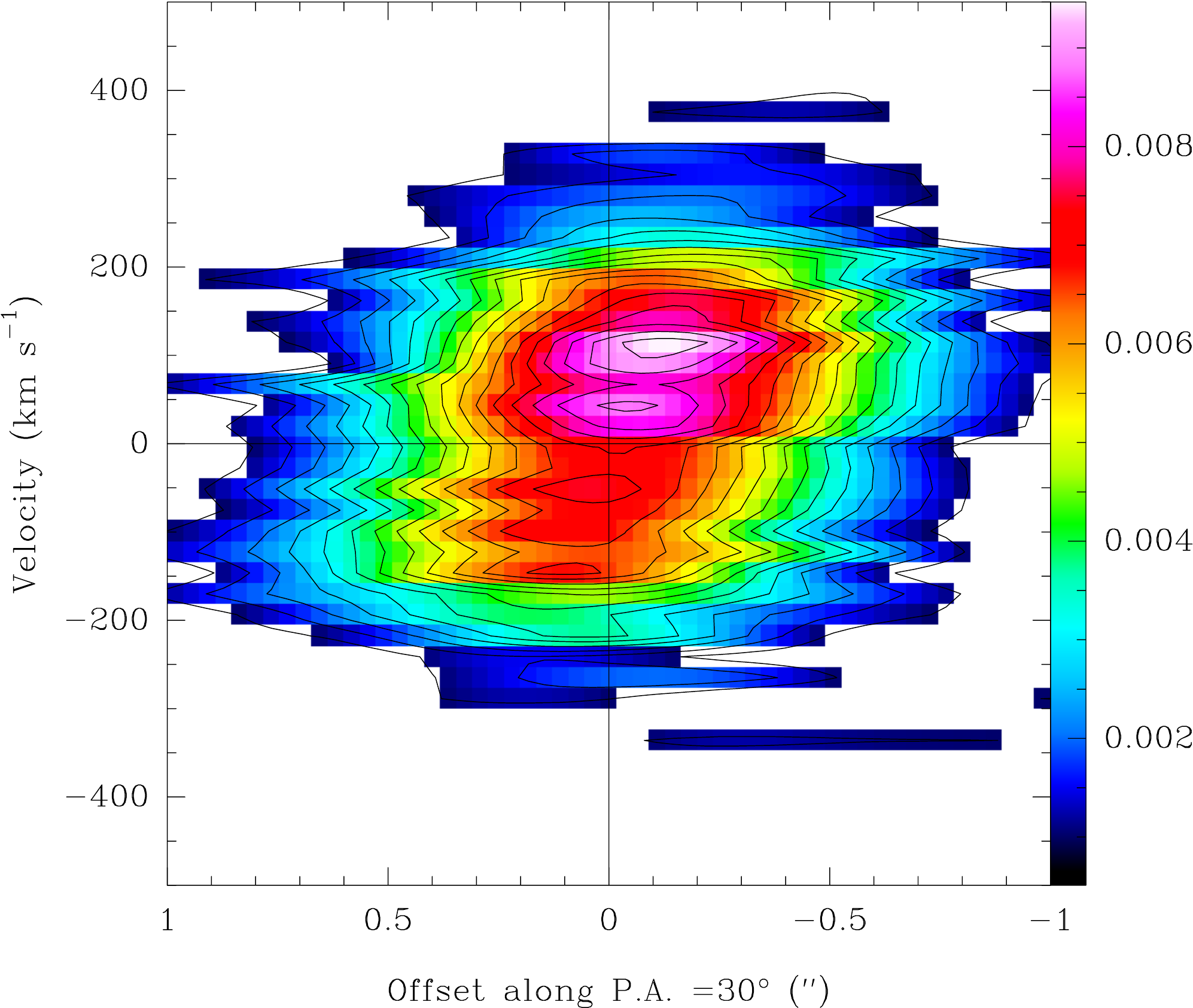}
    \includegraphics[width=8cm]{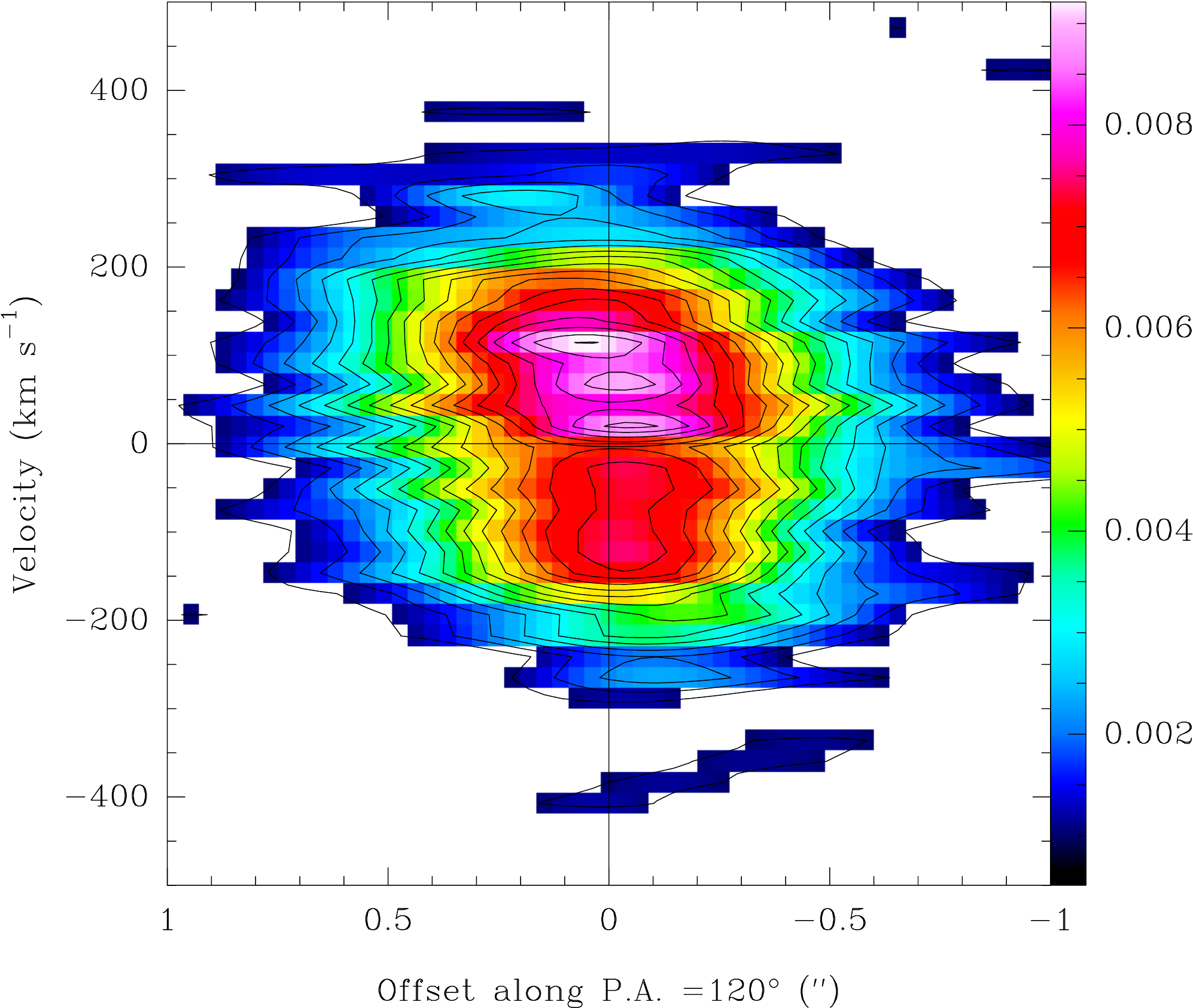}    
  \caption{PV diagrams of the continuum-subtracted [CII] emission along the line of nodes (PA= 30 deg, left panel), and orthogonal to it (PA=120 deg, right panel). Contours are 2 to 20$\sigma$, by $\sigma$, $\sigma= 0.14$ mJy/beam. Slices are extracted from a slit of 1.1 arcsec in width.  Zero offset position is (RA, DEC)=(23:10:38.90 ,18:55:19.82).}
              \label{pv-cii}
    \end{figure*}

  \begin{figure*}
   \centering
       \includegraphics[width=\textwidth]{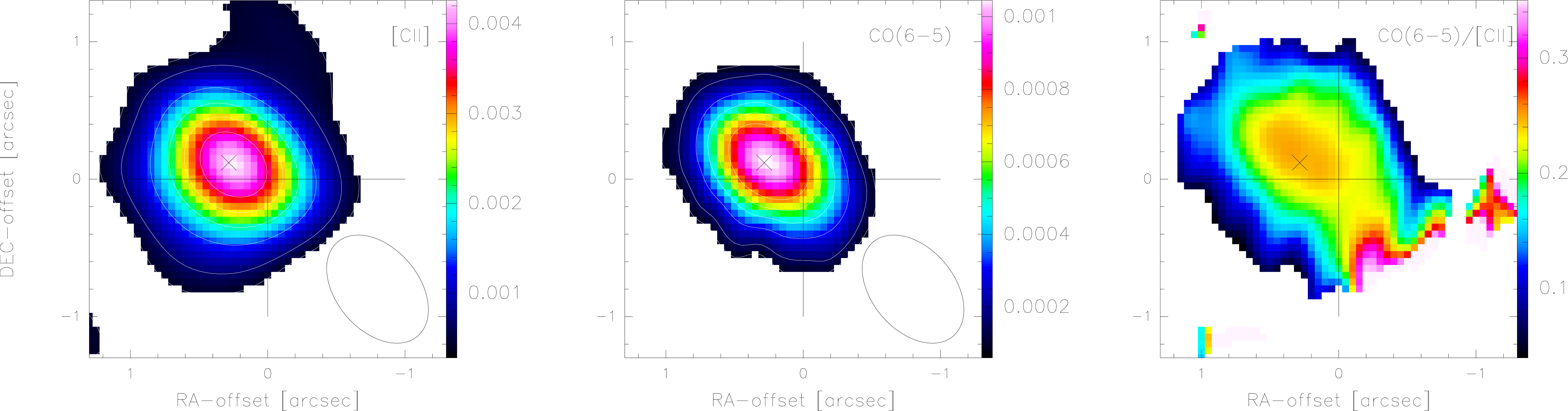}    
  \caption{[Left panel]: The [-500,500] km/s velocity-integrated [CII] map. The beam is $0.9\times 0.6$ arcsec$^2$, at  $\rm PA=49~ deg$. Levels are $-3,~ 3,~5,~ 10,~ 15,~ 20,~ 25\sigma,~\sigma=0.14 \rm ~ mJy/beam$. [Middle panel]: The velocity-integrated CO(6-5) map. The original angular resolution of CO data was degraded to match the same beam of [CII] data. Levels are $-3,~ 3,~5,~ 10,~ 15,~ 20,~ 25\sigma,~\sigma=27 \rm ~ \mu Jy/beam$. [Right panel]: The velocity-integrated CO(6-5)/[CII] ratio map. The small cross indicates (RA, DEC)=(23:10:38.90, 18:55:19.82). The large cross indicates the phase center.}
  \label{ratio}
    \end{figure*}

  \begin{figure}
   \centering
      \includegraphics[width=7cm]{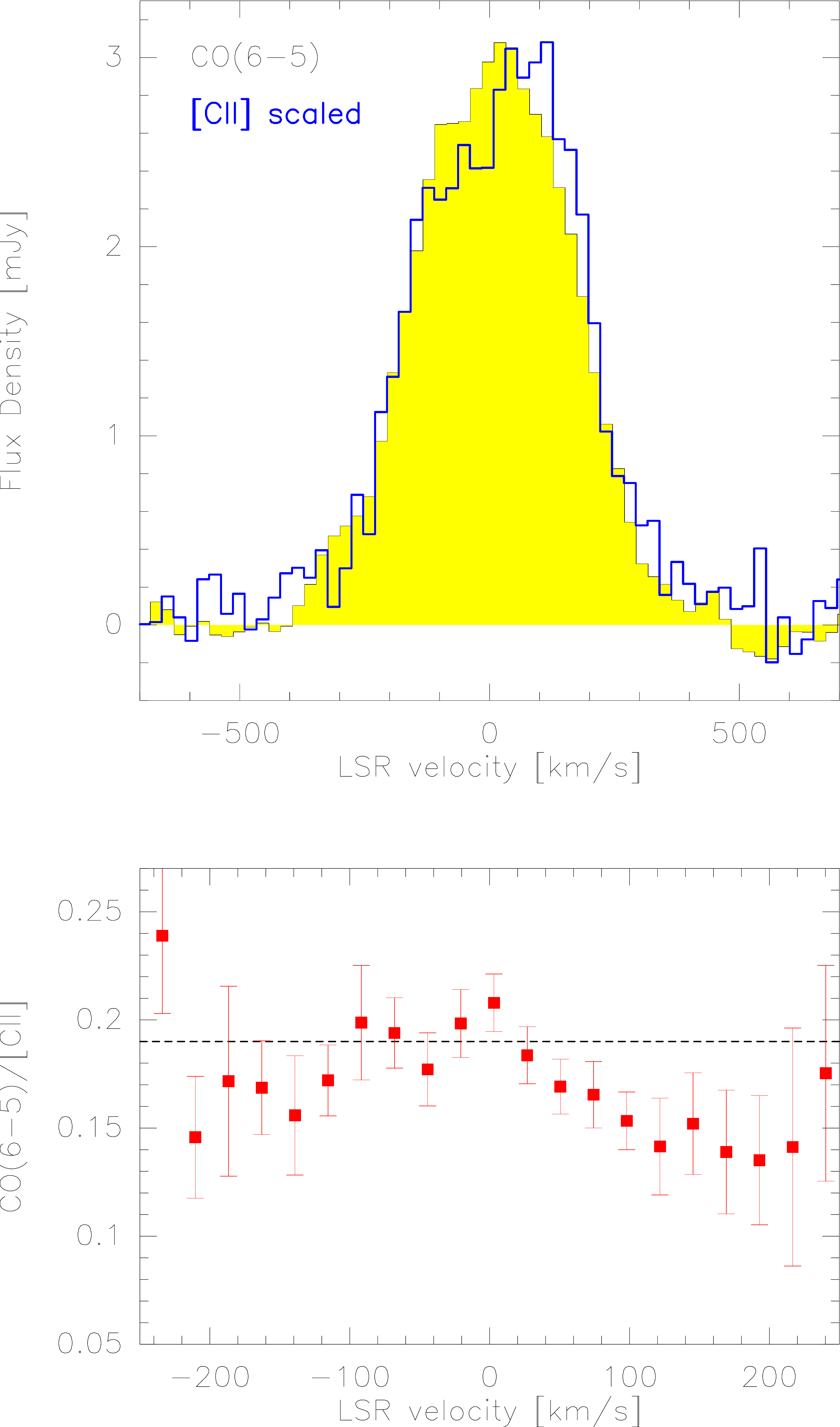}    
  \caption{[Upper panel]: The CO(6-5) (yellow filled histogram) and scaled [CII] (blue histogram) emission lines, extracted from 1 arcsec aperture centered at (RA, DEC)=(23:10:38.90, 18:55:19.82). 
Zero velocity has been set to the rest frame corresponding to $z=6.0025$ for both emission lines.
[Lower panel]: The ratio of the CO(6-5) and [CII] spectra in the velocity range $\rm [-250,+250]$ km/s. 
The dashed line indicates the velocity-averaged mean ratio. The binning is identical in the upper and lower panels.}
  \label{ratio-sp}
    \end{figure}

  \begin{figure}
   \centering
    \includegraphics[width=4.4cm]{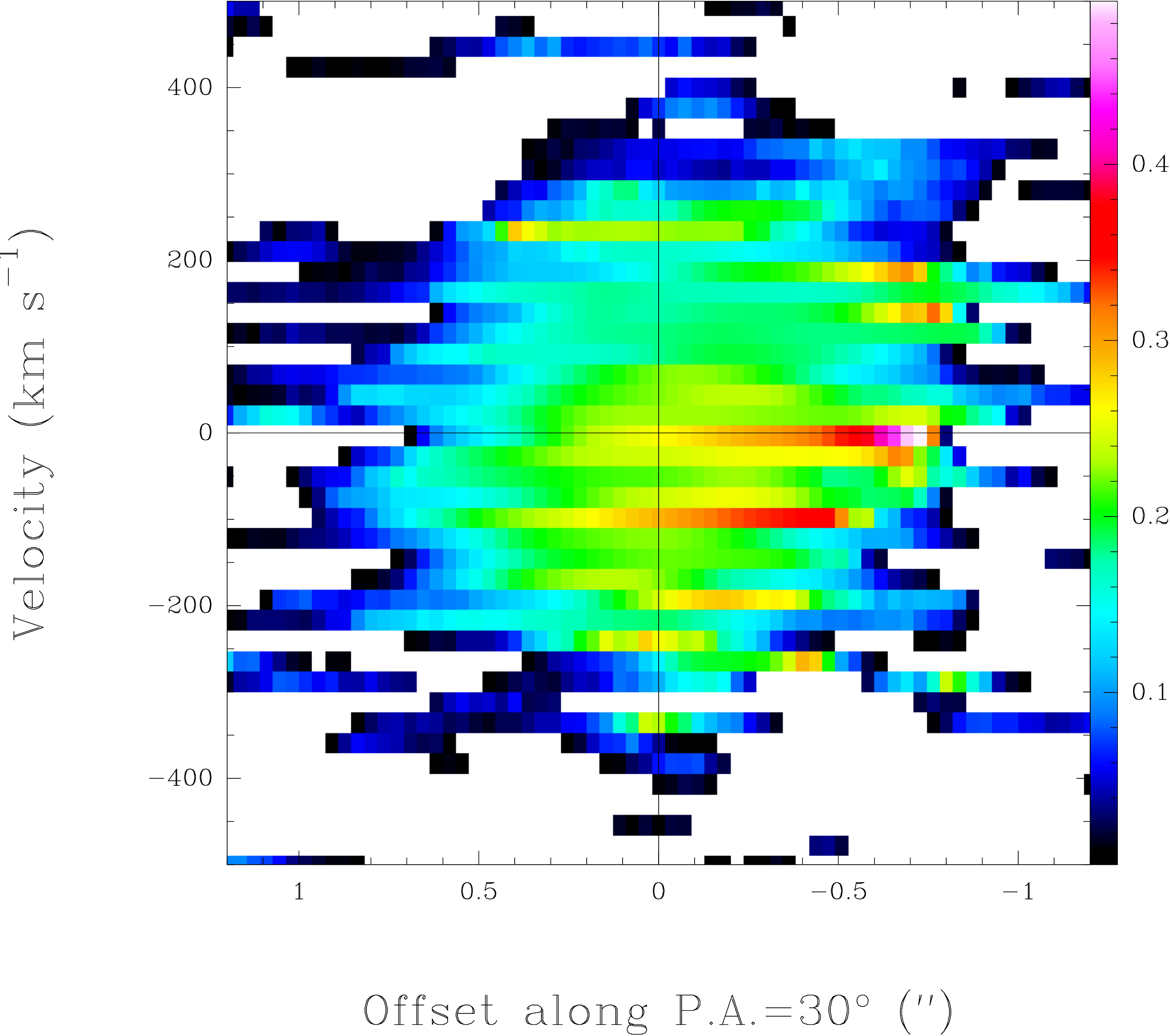}    
    \includegraphics[width=4.4cm]{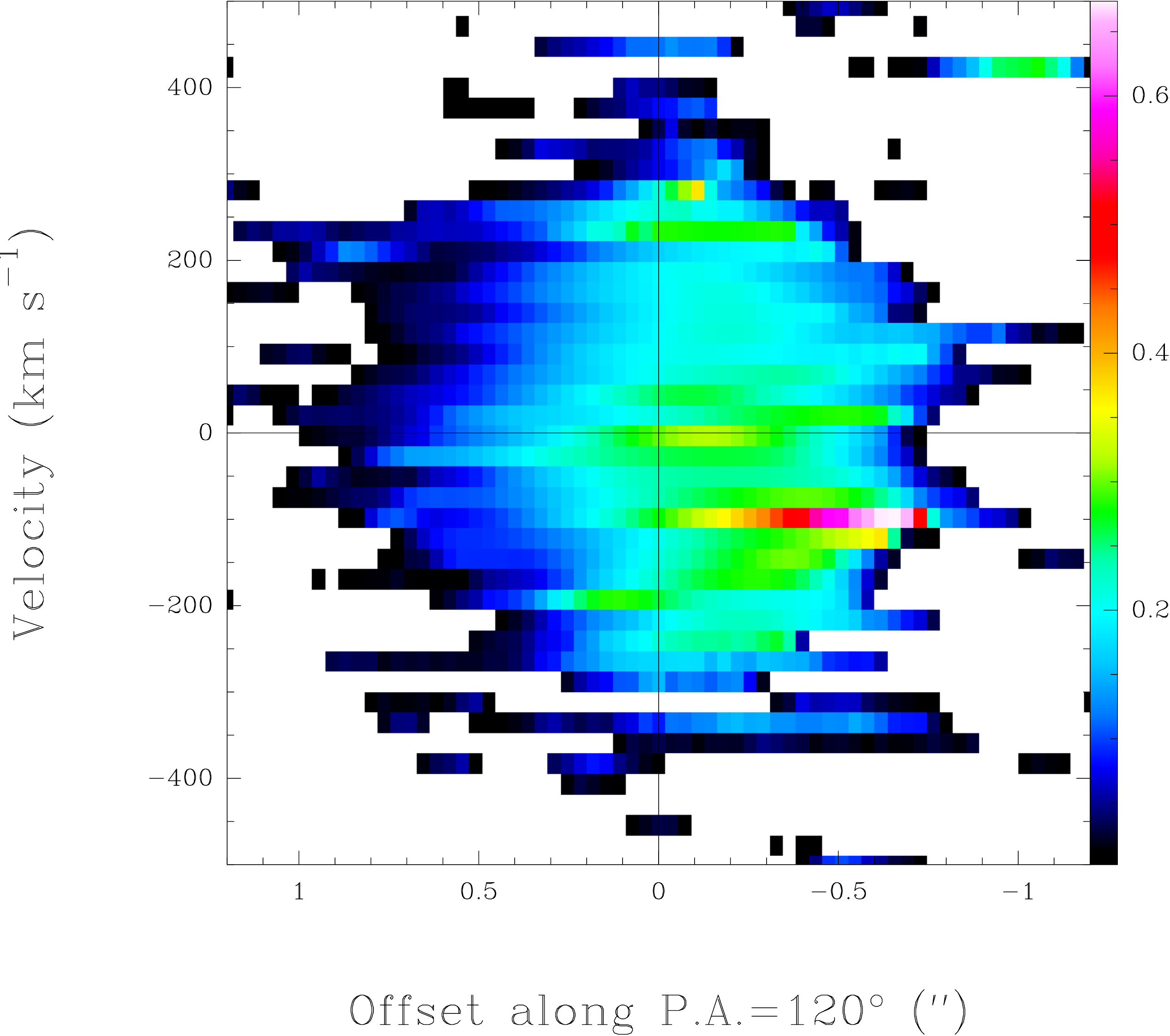}       
  \caption{The PV diagrams of the CO(6-5)/[CII] ratio along a $\rm PA = 30$ deg (left panel), and $\rm PA=120$ deg (right panel). Slit width is 0.9 arcsec. The original angular resolution of CO data was degraded to match that of the [CII] data.} 
              \label{pv-ratio}
    \end{figure}

\subsection{BH mass, gas mass, and dynamical mass}

We measured the monochromatic luminosity from the rest frame 1350 \AA~ continuum in the X-SHOOTER spectrum, $\lambda~L_\lambda(1350 \AA)= (9.8 \pm 1.9) \times 10^{45}$ erg s$^{-1}$.  
Mg~II  is not detected or very weak in the X-SHOOTER spectrum, probably due to the reduced sensitivity of the instrument at 1960 nm; we therefore estimate the BH mass based on C~IV $\lambda$1549\AA.  
 
We fit the CIV line with a single Gaussian component, from which we derive a FWHM=$11185\pm46$ km/s, and a $\chi^2=385.7$ for 181 degrees of freedom (d.o.f.). The C~IV line peak is located at 10727\AA. 
We note that, because it is probably affected by strong outflows, the line profile of C~IV is strongly asymmetric in high-luminosity QSOs, and it has been shown to over-predict BH masses, when adopting the Vestergaard \& Peterson (2000) correlation. 
The calibration by Coatman et al. (2016) provides a correction for these effects. By adopting the latter and 
a systemic redshift of $z=6.0025$, we derive a $\rm M_{BH} = (1.8\pm 0.5) \times 10^{9}~ M_{\odot}$.
This is a factor of approximately two smaller than the estimate by Jiang et al. (2016) based on Gemini data and on the Vestergaard \& Peterson (2000) correlation. The discrepancy is probably due to the different calibrations used.

  \begin{figure}
   \centering
              \includegraphics[width=\columnwidth]{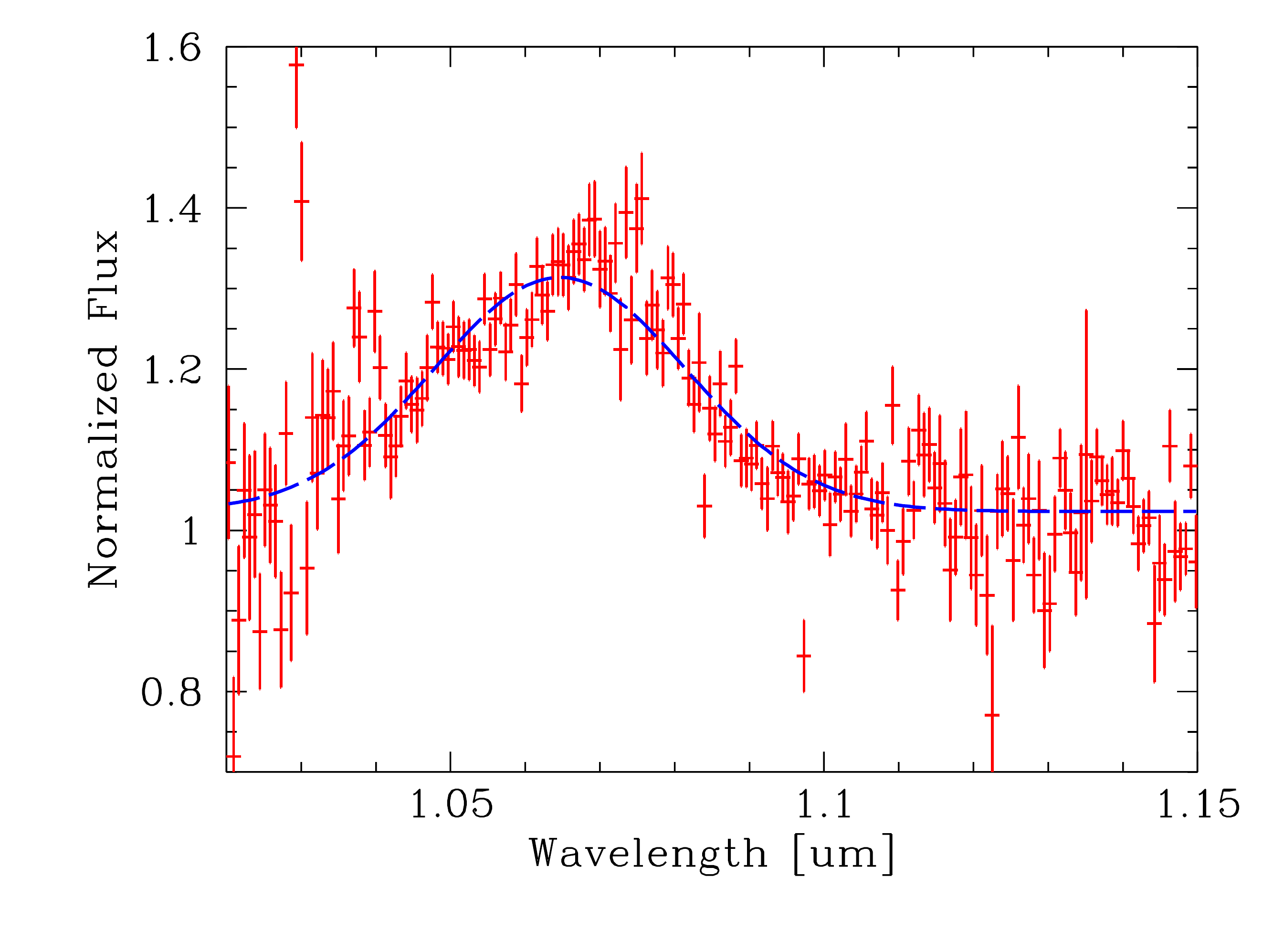}   
  \caption{The C~IV emission line from X-SHOOTER/VLT spectrum (red points). The blue dashed line shows the best fit with a Gaussian function.} 
              \label{civ}
    \end{figure}

Concerning the host galaxy mass budget, we can infer the molecular gas mass from the CO luminosity.
From the integrated intensity of CO(6-5), and using the relation $\rm L^\prime CO= 3.25\times 10^7 S_{CO}~\Delta v ~\nu^{-2}_{obs}~ D^2_L(1+z)^{-3}$ (Solomon \& Vanden Bout 2005), 
we derive a luminosity of $\rm L^\prime CO(6-5)=(4.3\pm0.1) \times 10^{10}~ K~ km/s ~pc^{-2}$. 
Stefan et al. (2015) and Wang et al. (2016) measured $\rm L^\prime CO (6-5) \sim 0.9~\times L^\prime CO (2-1)$ for the QSO J010013.02+280225.8 at $z=6.3$ and J114816.64+525150.3 at $z=6.4$, respectively. Since there is no measurement of lower J CO transition for J2310+1855, we adopt the same ratio. 
We assume thermalized emission (i.e., $\rm L^\prime~CO(2-1)=L^\prime~CO(1-0)$), and a conversion factor of $\alpha_{CO}=0.8~M_{\odot}~ \rm K^{-1}~ (km/s)^{-1} ~pc^{2}$ to derive the molecular gas mass. 
This conversion factor is currently thought to best represent QSO host galaxies (e.g., Carilli \& Walter 2013).
Under these assumptions, we find a molecular gas mass of $\rm M(H_2)=(3.2 \pm0.2) \times 10^{10}~(\alpha_{CO}/0.8)~~\rm M_{\odot}$. 
We note that using a Milky Way conversion factor ($4.3~M_{\odot}~ \rm K^{-1}~ (km/s)^{-1} ~pc^{2}$) would imply $\rm M(H_2)=(1.8 \pm0.2) \times 10^{11}~\rm M_{\odot}$.

Under the assumption that the observed CO velocity gradient is due to an inclined rotating disk, we derive the dynamical mass by applying the relation 
$\rm M_{dyn} \times sin^2(i) = 1.16~ 10^5 \times (0.75\times FWHM_{CO})^2 \times D$ (W13, Feruglio et al. 2014), where $\rm FWHM_{CO}=361\pm9$ km/s, and D is the source size in kiloparsecs (diameter). 
We adopt the definition of source size used by W13, equal to 1.5 times the FWHM size, $D=2.9$ kpc (i.e., full width at 20\% of the peak intensity for a Gaussian profile; Table 1).
We find $\rm M_{dyn} \times sin^2(i) = (2.4\pm0.5) \times 10^{10}~ M_{\odot}$. 
By using the virial relation $M_{dyn} =R~ v_{rot}^2 / G$ we find a similar value, $\rm M_{dyn} = 2.35 \times 10^{10}~ M_{\odot}$.
By applying the inclination derived from our data, we derive an inclination-corrected dynamical mass of $\rm M_{dyn} = 4.1^{+9.5}_{-0.5} \times 10^{10}~ M_{\odot}$. The upper limit $\rm M_{dyn}$ is derived from the lower limit inclination estimated from the minor/major axis ratio, $i\approx25$ deg.

\begin{table}
\caption{Derived properties of J2310+1855.} 
\label{table:1}      
\centering                          
\begin{tabular}{l l c }        
\hline
\hline
$z_{CO}$                                        &                                 & $6.0028\pm0.0003$  \\ 
L'CO(6-5)                               & $\rm [K~km/s~ pc^{-2}]$  & $(4.3\pm0.1)\times 10^{10}$ \\
M(H$_2$)                                        &  $[\rm M_\odot]$                &  $(3.2\pm0.2)\times 10^{10}$ \\
i                                               & [deg]                           &  $53$ \\
Mol. disk diameter              & [kpc]                   & $2.9\pm0.5$ \\
$\rm M_{dyn} sin^2(i)$          &  $\rm [M_\odot]$        & $(2.4\pm0.5)\times10^{10}$\\ 
$\rm M_{dyn}$$^{(a)} $                  &   $\rm [M_\odot]$               & $(4.1^{+9.5}_{-0.5})\times10^{10}$ \\
$\rm M_{BH}$$^{(b)}$            &  $\rm [M_\odot]$        & $(1.8\pm 0.5) \times10^9$        \\
$\rm M_{BH}/M_{dyn}$            &                                 &  $0.04^{+0.01}_{-0.035}$ \\ 
$\rm L_{bol}^{(c)}$                             & $\rm [L_{\odot}$]               & $9.3\times10^{13}$ \\ 
$\dot M_{acc}$                          & $\rm [M_\odot/yr]$         & $63\pm11$ \\
$L_{FIR}$                                       & $\rm [L_{\odot}]$               & $(1.70\pm0.18)\times10^{13}$ \\
SFR     $^{(d)}$                                &  $\rm [M_\odot/yr]$     & $1250\pm900$ \\
$\rm M(H_2)/M^*$                                &                                 & $\approx4.4$ \\ 
$\rm M(BH)/M^*$                                 &                                 & $\approx0.25$ \\ 
\hline
\end{tabular}
\tablefoot{ The upper limit $\rm M_{dyn}$ is (a)   derived using the lower limit inclination $i_{min}=25$ deg,
$ (b)$ derived from C~IV FWHM and the 1350\AA~ continuum luminosity, 
$(c)$ derived from rest-frame 1450\AA~ magnitude and assuming a bolometric correction $L_{bol}=4.2~ L_{1450}$ (Runnoe et al. 2012a,b) and 
$(d)$ derived from $\rm L_{FIR}$ (from W13) and the Kennicutt \& Evans (2012) relation.}
\end{table}

\subsection{Search for other line or continuum emitters.}

We scanned all spectral windows of the band 3 and band 6 data cubes in space and frequency, searching for continuum or line emitters. To this purpose, we degraded the original spectral resolution to 47.4 km/s.
Candidate detections are considered if signal is detected in at least three contiguous spectral channels with a global (i.e., velocity integrated) statistical significance of $\geq4.5\sigma$ in the velocity-integrated map.

\begin{figure}
   \centering
    \includegraphics[width=4cm]{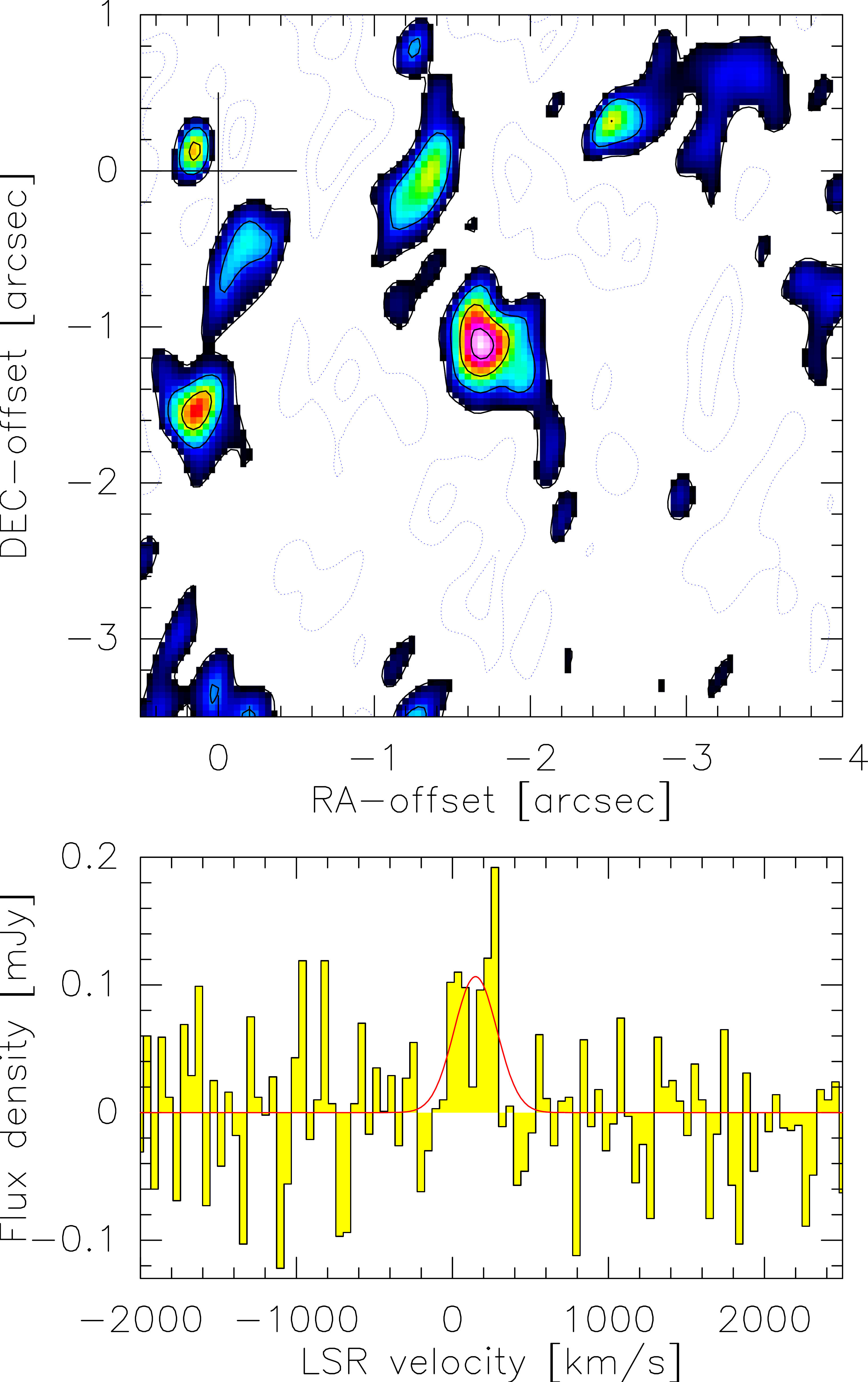}        
            \includegraphics[width=3.95cm]{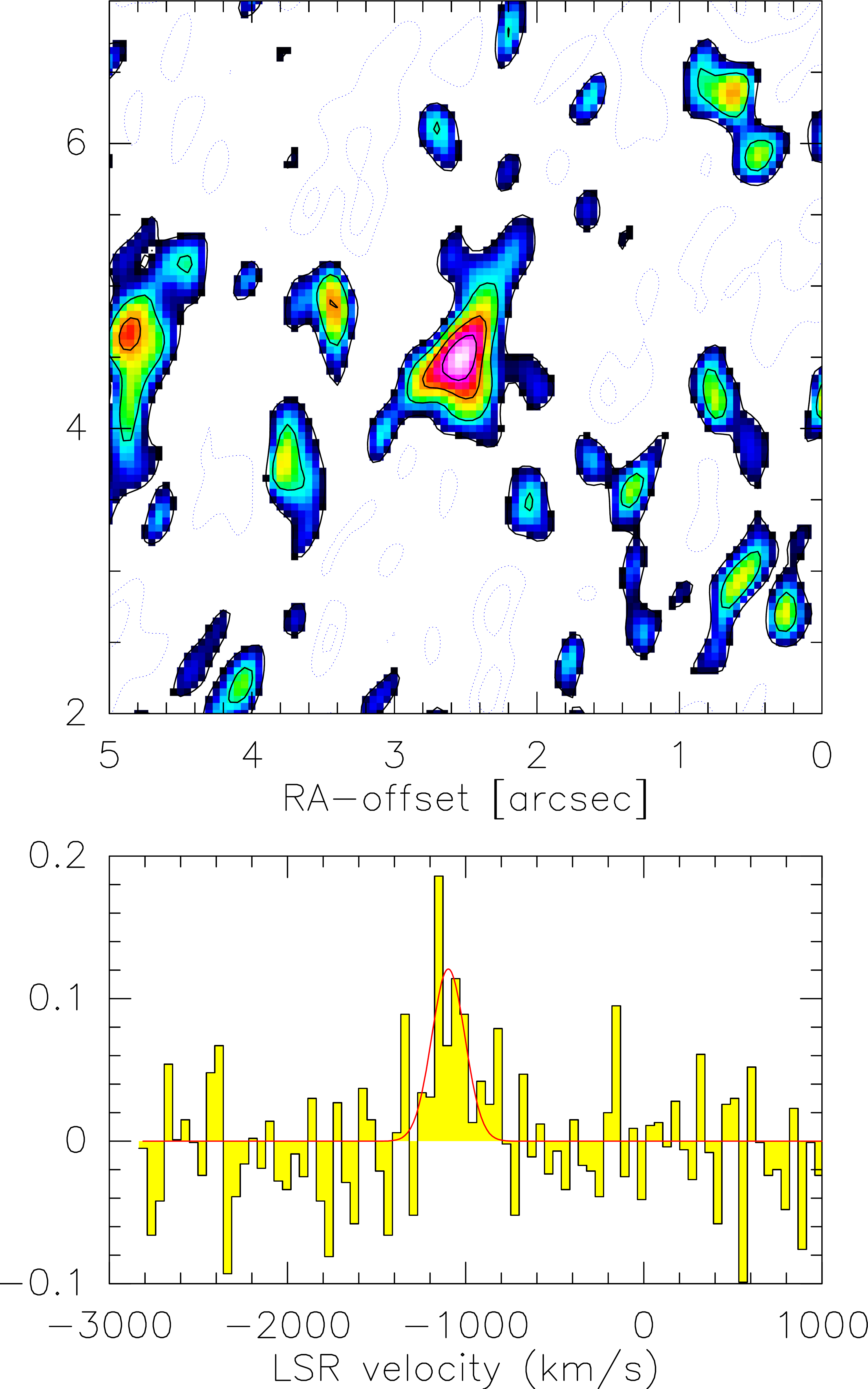}       
  \caption{The velocity-integrated maps (upper panels) and the spectra (lower panels) of the candidate line emitters detected in band 3. In the upper-left panel the CO emission from the QSO host galaxy has been fitted and subtracted (the cross indicates the phase center).  
The spectra were extracted using masks enclosing the $\geq2\sigma$ level on the velocity-integrated maps. 
Contour levels are $-4,-3,-2,-1,1,2,3,4\times \sigma$, $\sigma=0.035, 0.059~ \rm mJy~ beam^{-1}$ in the upper--left and upper--right panels, respectively.
} 
              \label{galaxies}
    \end{figure}

In the following we describe the candidate emitters selected from this analysis. 
One line emitter, detected in the band-3 data with a statistical significance of 5$\sigma$ and with prior information of its redshift, is presented in D'Odorico et al. (2018).  
Other two line emitters are detected in band-3. Figure \ref{galaxies} shows the velocity-integrated line maps of these two candidate sources, which are detected with a significance of $4.5\sigma$ in the band-3 data.
The line peaks are detected at velocity $\sim200$ and $\sim1100$ km/s from the CO(6-5) of the QSO host galaxy.  
Their positions are [RA, DEC]=[23:10:38.769,18:55:18.49], [RA, DEC]=[23:10:39.068,18:55:24.16], respectively.  
Because the first candidate has a frequency similar to the emission line in the QSO, the possibility that this is due to an artifact (e.g., cleaning residual) cannot be ruled out. 
By fitting the lines with a Gaussian function we find $\rm FWHM = 300\pm100$ and $220\pm60$ km/s, 
and integrated flux densities of $0.036\pm0.008$ Jy km/s and $0.032\pm 0.007$ Jy km/s, respectively. 
The underlying 3 mm continuum is undetected in both these cases.  
At their position we do not detect any continuum or line in band 6 either.
 
A secure identification of these candidate lines requires the detection of at least another emission line. 
Here we attempt to assess whether these candidate emission lines are genuine by comparing our findings with the results of the ASPECS Survey of the Hubble Ultra-Deep field  (Walter et al. (2016), Decarli et al. 2016). 
In their work, they provide molecular emission line number counts, and the CO luminosity function based on their ALMA observation covering the band 3 with a total bandwidht of 30.75 GHz, and with a sensitivity comparable to ours (0.17 mJy/beam per 23.7 km/s around 95 GHz, with the same field of view).  
According to their number counts of molecular emission lines, and to their emission line luminosity function, 
and by rescaling for our narrower bandwidth (3.38 GHz versus the $\sim30.75$ GHz covered by their survey), 
0.2-0.8 CO-emitting sources with flux density in the range 0.03-0.04 Jy km/s are expected in the area covered by one ALMA antenna beam at 3 mm.   
We therefore find a factor of $\sim 3-15$ excess in the number of line-emitting sources in our data compared to the expectation based on ASPECS. If we considered as genuine only the source detected with the highest 
statistical significance (D'Odorico et al. 2018), we would still be at the upper boundary of the expected line counts.
We note, however, that the  $log N -- log S$ of molecular lines is poorly known, and that the estimates based on the ASPECS survey have very large uncertainty. Their estimates are also based on a single field (one ALMA pointing at 3 mm), and therefore do not take into account the cosmic variance. 
In this context, the UDF can be considered a random field, whereas our observation targets the field around a rare luminous QSO, and therefore likely probes a biased overdensity region (Decarli et al. 2016, Bischetti et al. 2018).

  \begin{figure*}
   \centering
  \includegraphics[width=6cm]{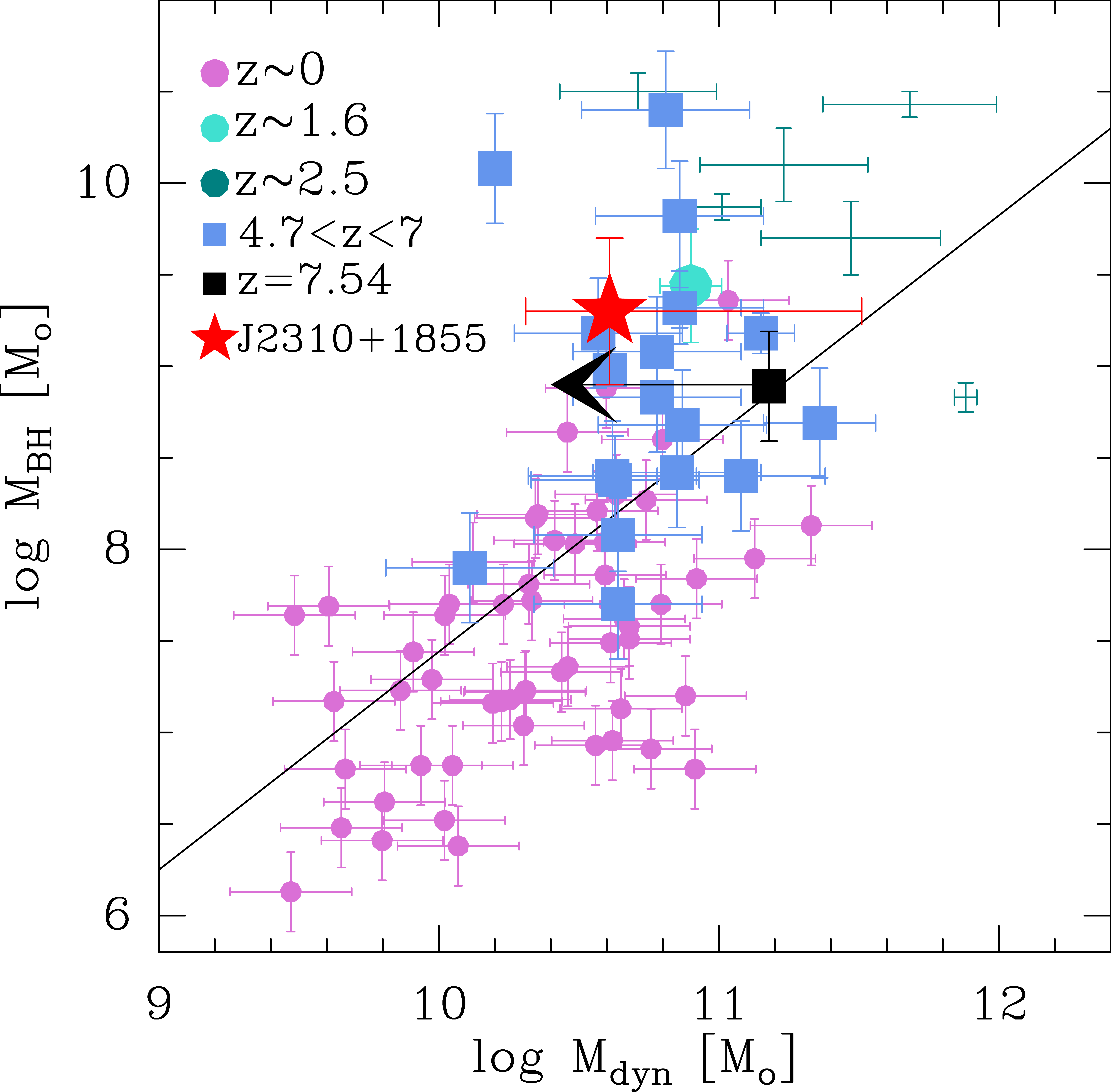}
   \includegraphics[width=6cm]{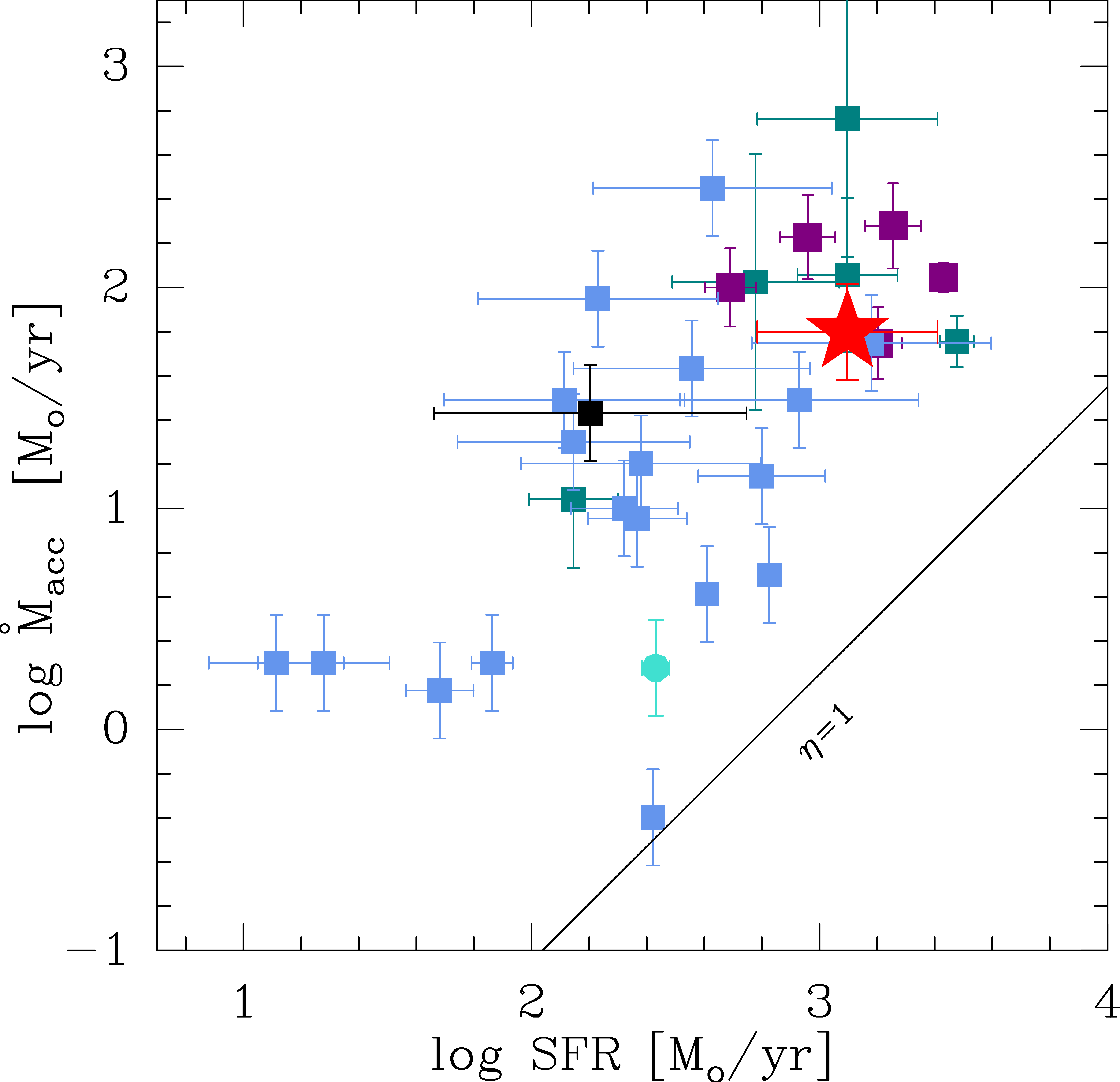}
    \includegraphics[width=6cm]{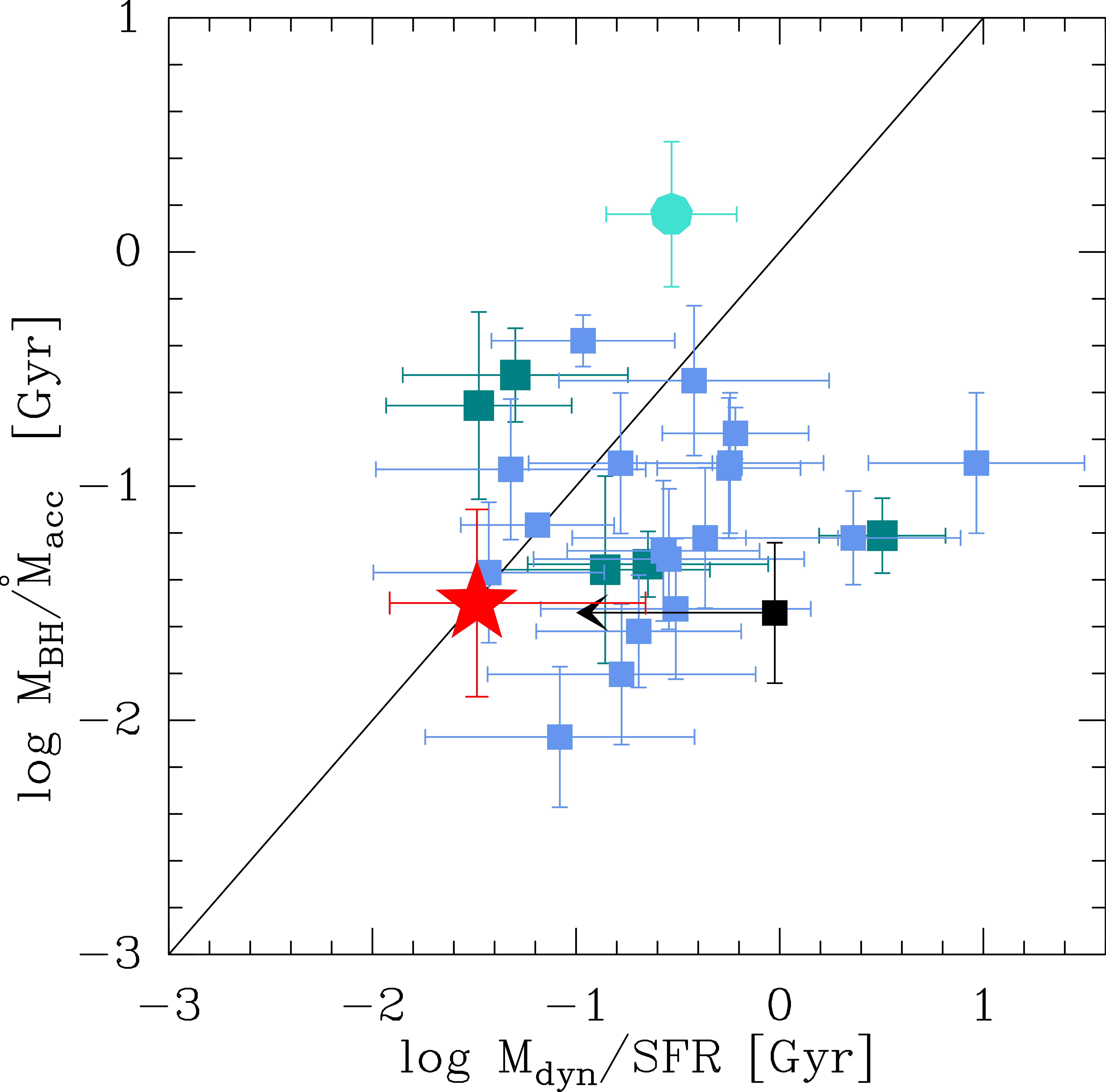}  
  \caption{[Left panel]: $\rm M_{BH}$ vs. $\rm M_{dyn}$ for J2310+1855 (red star) and a compilation of AGNs.  
Violet circles: IBISCO hard X-ray selected sample at z$<0.05$ (Feruglio et al. in prep). Cyan circle: Brusa et al. (2017). 
Green symbols: red QSOs at $z \sim2.5$, for these we plot the ranges given in Banerji et al. (2012, 2017). Blue squares: QSO at $z>4.7$ from Willott et al. (2013, 2015, 2017), Kimball et al. (2015), Wang et al. (2016), De Rosa et al. (2014), Venemans et al.  (2016, 2017a), Trakhtenbrot et al. (2017), Decarli et al. (2018). Black symbol: $z=7.54$ QSO from Venemans et al. (2017b), Ba\~nados et al. (2018). For the QSO for which there is no estimate of the inclination available, we have assumed $i=50\deg$ (see e.g., Decarli et al. 2018). The solid line is the local $\rm M_{BH}-M_{bulge}$ relationship (Kormendy \& Ho 2013).
[Middle panel]: The BH accretion rate, $\rm \dot M_{acc}$, vs. the SFR. Same symbols as in the left panel, except purple squares = hyper-luminous QSO at $z=3-4$ from the WISSH sample (Bischetti et al. 2017, Duras et al. 2017, Vietri et al. 2018). The line is the locus of $\rm \eta=\dot M_{outflow}/SFR=1$, assuming the scaling relation of $\dot M_{outflow}$ with $L_{bol}$ derived for molecular outflows (Fiore et al. 2017).
[Right panel]: Black hole growth timescale, $\rm M_{BH}/\dot M_{acc}$ vs. the galaxy growth time, $\rm M_{dyn}/SFR$. The solid line is the 1:1 relation.
}
              \label{mbh-mdyn}
    \end{figure*}

\section{Discussion}

We mapped and resolved the CO(6-5) line emission in the host galaxy of the QSO J2310+1855. 
We measured a molecular gas mass of $\rm M(H_2)=(3.2 \pm0.2) \times 10^{10}~(\alpha_{CO}/0.8)~\rm M_{\odot}$, and a size of the CO-emitting region of 2.9 kpc. 
Under the assumption that the observed velocity gradient is due to a rotating disk, we derived 
 an inclination of $i=53$ deg from the minor/major axis ratio, and a dynamical mass of $\rm M_{dyn} = 4.1 \times 10^{10}~ M_{\odot}$ within the inner 2.9 kpc region, which is a factor of approximately two  smaller than the W13 value.  
For the lower limit inclination implied to our data, $i=25$ deg, the dynamical mass would be $\rm M_{dyn} = 1.3 \times 10^{11}~ M_{\odot}$.

We note that the estimate of the dynamical mass depends linearly on the disk size and on $sin^2(i)$, which in turn depend on the signal-to-noise ratio (S/N) of the data and on the angular resolution. Therefore data with a high S/N and good angular resolution are essential elements to derive accurate dynamical masses.  
The discrepancy between our estimate and the W13 may be due to the smaller source size that we measured based on CO data ($(0.33\pm0.06) \times (0.20\pm0.04)$ arcsec vs. the  $(0.56\pm0.03) \times (0.39\pm0.04)$ arcsec by W13), and to the slightly larger inclination. 

We find that most of the mass within a region of 2.9 kpc in  diameter is in the form of dense molecular gas. 
By assuming that the stellar mass is $\rm M^*\sim M_{dyn}-M(H_2)-M(BH)=7.2\times 10^9~M_\odot$, we derive a molecular gas fraction of $\mu=M(H_2)/M^*\approx4.4$. 
For comparison, the molecular gas fraction in z=3-4 main sequence galaxies is at most $\mu=2.5$ (Tacconi et al. 2018, Genzel et al. 2015).  
The BH to stellar mass ratio is $\rm M(BH)/M_*\approx 0.25$, which is very large. We note that this value refers to the inner 2.9 kpc of the host galaxy only.

The ratio $v_{rot}/\sigma=1-2$ indicates that the molecular gas is turbulent, similarly to what is found in z=2 star-forming galaxies by Tacconi et al. (2013) (and references therein), and in z=3 star-forming galaxies by Gnerucci et al. (2011) and Williams et al. (2014). 
The turbulence may be due to a thick, dynamically hot disk, and/or to outflows/inflows. The current observations do not allow us to discriminate between these two possibilities. 

We derived the Toomre parameter $Q= a~ \sigma /f_{gas}~ v_{rot}$, where $a=1.4$ for flat rotation curves ($a=1$ for keplerian disks), and $f_{gas}$ is the gas fraction (Genzel et al. 2014). 
We find $Q\sim0.2-0.5$, meaning that cloud fragmentation likely occurs in the disk. 
We note, however, that small values of Q do not necessarily imply local gravitational instability (Romeo \& Agertz 2014, Romeo et al. 2010).
We caution that the estimated Q is an average value, does not exclude local variations through the disk, and is  strongly dependent on angular resolution.

We studied the CO/[CII] ratio by comparing the CO and [CII] line profiles and their maps. 
We find that the CO and [CII] lines show similar profiles. 
Our analysis hints at a broader velocity distribution of the [CII]-emitting gas than the CO(6-5)-emitting gas, within $\pm250$ km/s from the CO line peak (Fig. \ref{ratio-sp}, \ref{pv-ratio}), suggesting that  there may be differences in the kinematics and spatial distributions of dense clumps of molecular material traced by CO(6-5) ($n_{crit}\sim 10^6\rm cm^{-3}$), and the low-ionization gas traced by [CII] ($n\sim 10^3~\rm cm^{-3}$). 
Additional observations are required to assess this finding.
We found a mean $\rm CO(6-5)/[CII]$  of $\sim 0.19$, while a lower value, of $\sim 0.05$, is expected from emission models of high-z galaxies that do not host a QSO and whose ionization field is due only to the stellar component (Pallottini et al. 2017, Vallini et al. 2018).  
The larger [CII] extension may imply that some of the [CII] emission comes from the diffuse and less dense ISM. 
The higher CO(6-5) concentration might also result from the presence of the QSO intense UV field, boosting high J CO lines. 
We note, however, that the bulk of the [CII] emission from these bright QSOs at $z\sim6$ comes from the outer layers of photodissociation regions (PDR), and that a high CO(6-5)/[CII] ratio can also be explained within classical PDR modeling, assuming $\rm n_H>3~ 10^5$ cm$^{-3}$, and a radiation field intensity of $< 1000~ G_0$, or $\rm n_H>16~ \rm cm^{-3}$ and higher radiation fields ($>1000~ G_0$).

In the following we use $\rm M_{dyn}$ estimate derived from CO(6-5), which traces the densest regions of molecular clouds, where most of the SF should occur, and our new estimate of the BH mass, $\rm M_{BH} = (1.8\pm 0.5) \times 10^{9}~ M_{\odot}$. 
Figure \ref{mbh-mdyn}, left panel, shows the BH mass versus the dynamical mass for J2310+1855 and a compilation of AGN and QSO for which both the BH mass and the dynamical mass has been measured, the latter through [CII] or CO emission lines (see caption of Fig. \ref{mbh-mdyn} for details about the literature samples). 
 In this plot we compare individual measurements of $\rm M_{dyn}$ with the local $\rm M_{BH}-M_{bulge}$ relationship (Kormendy \& Ho 2013).
For J2310+1855 we find a mass ratio $\rm M_{BH}/M_{dyn}=0.04^{+0.01}_{-0.035}$.  
If the molecular disk is not seen at a very small inclination, the dynamical mass derived here places J2310+1855 significantly above the local $\rm M_{BH}-M_{bulge}$ correlation, similarly to other z=5-7 QSO.

From the QSO bolometric  luminosity we derive the BH accretion rate, $\rm \dot M_{acc}=L_{bol}/\epsilon c^2 $, assuming $\epsilon=0.1$. The middle panel of 
Figure \ref{mbh-mdyn}  shows $\dot M_{acc}$ versus the SFR of the host galaxy.
Here we also plot hyper-luminous QSOs at $z=3-4$ from the WISSH sample (Bischetti et al. 2017, Duras et al. 2017, Vietri et al. 2018, note that these do not have a measured $\rm M_{dyn}$ and therefore they are not shown the left panel). 
For QSOs with $\rm L_{bol}=10^{47}$ erg/s, Duras et al. (2017) found that about 50\% of the FIR luminosity is due to reprocessed radiation from the AGN, therefore we corrected SFRs by this amount in the relevant $L_{bol}$ range. 
We plot the line corresponding to the mass loading factor of molecular winds equal to unity, $\rm \eta=\dot M_{outflow}/SFR =1$. 
This was derived from the scaling relation of $\rm \dot M_{outflow}$ with $L_{bol}$, $\rm \log \dot M_{outflow} = 0.76 \log L_{bol} -32$ (Fiore et al. 2017).  
Because $\rm L_{bol}= \epsilon \dot M_{acc} c^2$, and assuming $\rm \epsilon=0.1$, according to the scaling relation of Fiore et al. (2017), 
$\rm \log \dot M_{acc} = 1.3~ log(SFR) - 3.65$, for $\eta=1$.  
We find that, if the scaling relations hold at these extreme regimes and at high redshift, all the high-redshift QSOs, including J2310+1855, are in a regime where the QSO can drive massive outflows with $\rm \eta>1$. This suggest that appropriately sensitive observations have the potential to reveal molecular outflows in these QSOs (see e.g., Brusa et al. 2017). 

$\rm \dot M_{acc}$ versus the SFR are the derivative quantities of BH mass and dynamical mass. Therefore, in principle, they are useful to understand how BH and galaxy grow to reach their location on the left-hand diagram of Fig. 11, if the sources are caught in the phase when most of both the stellar and the BH mass is being assembled.
This is realistic at $z\sim4-6$, but likely not the case at low redshift, therefore we omit the $z\sim 0 $ sample in this plot.
To verify this, we compare the growth timescale of the BH at the current accretion rate ($\rm M_{BH}/\dot M_{acc}$) with the growth timescale of the host galaxy at the current SFR ( $\rm M_{dyn}/SFR$). 
The right panel of Figure \ref{mbh-mdyn} shows that the BH growth time is similar to the galaxy growth timescale if we consider the dynamical mass as a proxy for the stellar mass. 

Concerning other continuum or line emitters, we detected two candidate emission lines within $\sim 1000$ km/s of the QSO redshift in the band-3 data, located at projected distances of 12 and 29 kpc from the QSO position.  
A secure identification of these candidate sources requires the detection of at least one other emission line. 
If the candidate lines were genuine, and if we identified them with CO(6-5), and under the same assumptions used for deriving the molecular gas mass of the QSO, these would correspond to molecular gas reservoirs of $\rm M(H_2)\approx (0.8- 1) \times 10^{9}~(\alpha_{CO}/0.8)$ M$_\odot$.   
If we apply the calibration of Greve et al. (2014) between the total IR luminosity and $\rm L^\prime CO(6-5)$, and the Kennicutt (1998) relation corrected for a Kroupa (2011) IMF, these would convert into SFRs $\approx 50$ $\rm M_{\odot}/yr$. 
This is a very rough estimate, but is consistent with the nondetection of these sources in 200 $\mu$m continuum in band 6. In fact, by assuming a typical SED of a star-forming galaxy with dust temperature in the range $\rm T_{dust}\sim 30-50~ K$, we would expect a $3\sigma$ detection of the continuum for a $\rm SFR=200 ~M_{\odot}/yr$ for a galaxy located at $ z\sim6$.
If confirmed, these line emitters, together with the galaxy presented in D'Odorico et al. (2018), would trace an overdensity of galaxies located close to the QSO, in a configuration similar to that found for other high -z QSOs (e.g., Decarli et al. 2017), and at $z\sim3$ (Fogasy et al. 2017, Bischetti et al. 2018). 
We argue that the physical separations of these galaxies from the QSO are typical scales of the circum-galactic medium (CGM), therefore these galaxies may eventually be able to merge with the QSO host, thus contributing to the growth of a giant elliptical galaxy, in agreement with most models of hierarchical galaxy formation. Without confirmation of the two candidate detections presented here however, we are disinclined to make any firm conclusions on the close environment of this QSO.

\section{Conclusions}

Using ALMA we mapped the CO(6-5) and [CII] emission lines and the sub-millimeter continuum of the z$\sim6$ QSO SDSS J231038.88+185519.7. The angular resolution and sensitivity of our data allowed us to resolve the dense molecular gas emission in the host galaxy of the QSO.
Our findings are summarized below.

 We measure a molecular gas mass of the QSO host galaxy of $\rm M(H_2)=(3.2 \pm0.2) \times 10^{10}\rm M_{\odot}$, 
a size of the molecular disk of $2.9\pm0.5$ kpc, and an inclination of $i\approx53$ deg. We derived a dynamical mass of $\rm M_{dyn} = 4.1 \times 10^{10}~ M_{\odot}$, and a molecular gas fraction $\mu=M(H_2)/M_*\approx 4.4$, which is larger than that found for main sequence galaxies at z=3-4. 

We derive a ratio $v_{rot}/\sigma=1-2$, suggesting high gas turbulence and/or outflows/inflows, and a Toomre parameter $Q\sim 0.2-0.5$, indicating cloud fragmentation.

We provide a new estimate of the BH mass based on C~IV emission line detected in the X-SHOOTER/VLT spectrum, $\rm M_{BH} = (1.8\pm 0.5) \times 10^{9}~ M_{\odot}$. 
The dynamical mass derived here places J2310+1855 above the local $\rm M_{BH}-M_{bulge}$ correlation, similarly to other high-z QSOs. 
We find that for J2310+1855 and for most QSOs, the current BH growth rate is similar to that of its host galaxy. 
We argue that all the high-redshift QSOs, including J2310+1855, are in a regime where the QSO can drive massive outflows with loading factors $\eta>1$, if the AGN wind scaling relations hold at these extreme regimes and at high redshift. 

We compare the CO(6-5) to the [CII] emission, finding that they have similar line profiles. 
 The observed slight discrepancies in the line profiles of CO(6-5) and [CII] near the line peaks, and the hints of a broader velocity distribution of the [CII]-emitting gas compared to the CO(6-5), require additional observations to be confirmed.

We detect two candidate emission lines within $\sim 1000$ km/s of the QSO CO line.  
With the data used here, it is difficult to determine whether the two detections presented in this work are genuine. These candidate lines require confirmation by the detection of at least one additional emission line.
If they were genuine, 
given their close projected distances from the QSO, these galaxies, together with the CO(6-5) emitter detected in the proximity of the QSO (D'Odorico et al. 2018), would trace an overdensity around the QSO, and may eventually be able to merge with the QSO host galaxy, thus contributing to the hierarchical growth of a giant elliptical galaxy, in agreement with most models of galaxy formation.

\begin{acknowledgements}
We thank the referee for his thorough review, and for his highly appreciated comments and suggestions. 
This paper makes use of the following ALMA data: ADS/JAO.ALMA\#2015.1.00584.S and 2015.1.00997.S. ALMA is a partnership of ESO (representing its member states), NSF (USA) and NINS (Japan), together with NRC (Canada), MOST and ASIAA (Taiwan), and KASI (Republic of Korea), in cooperation with the Republic of Chile. The Joint ALMA Observatory is operated by ESO, AUI/NRAO and NAOJ.
Based on observations made with ESO Telescopes at the La Silla Paranal Observatory under programme ID 098.B-0537(A).
CF acknowledges support from  the European Union Horizon 2020 research and innovation programme under the Marie Sklodowska-Curie grant agreement No 664931. 
FF acknowledges financial support from INAF under the contract PRIN-INAF-2016 {\it FORECAST}. 
R.M. and S.C. acknowledge support by the Science and Technology Facilities Council (STFC). 
R.M. acknowledges ERC Advanced Grant 695671 {\it QUENCH}. 
EP acknowledges financial support from INAF under the contract PRIN-INAF-2012. 
LZ acknowledges financial support under ASI/INAF contract I/037/12/0.

\end{acknowledgements}

\end{document}